\documentclass[twocolumn,showpacs,aps,amsmath,amssymb,prl,superscriptaddress]{revtex4-1}   %showpacs,
\usepackage{amsfonts}
\usepackage{bm}
\usepackage{setspace}
\usepackage{graphicx}
\usepackage{color}
\usepackage{multirow}
\usepackage{verbatim}
\usepackage{float}

\newcommand{\be}{\begin{equation}}
\newcommand{\ee}{\end{equation}}
\newcommand{\bea}{\begin{eqnarray}}
\newcommand{\eea}{\end{eqnarray}}
\newcommand{\bsube}{\begin{subequations}}
\newcommand{\esube}{\end{subequations}}
\newcommand{\Eq}[1]{Eq.~(\ref{#1})}
\newcommand{\Eqs}[1]{Eqs.~(\ref{#1})}
\newcommand{\Fig}[1]{Fig.~\ref{#1}}

\newcommand{\br}{\mathbf{r}}

\newcommand{\ep}{\epsilon}

\newcommand{\cl}[1]{\textcolor{blue}{#1}}
\begin{document}

\title{Localized Orbital Scaling Correction for Systematic Elimination of Delocalization Error in Density Functional Approximations}

\author{Chen Li}
\affiliation{Department of Chemistry, Duke University, Durham, North Carolina 27708, USA}

%\author{DaDi Zhang}
%\affiliation{Synergetic Innovation Center of Quantum Information and Quantum Physics, Hefei National Laboratory for Physical Sciences at the Microscale, University of Science and Technology of China, Hefei, Anhui
%230026, China}

\author{Xiao Zheng} %\email{xz58@ustc.edu.cn}
\affiliation{Synergetic Innovation Center of Quantum Information and Quantum Physics, Hefei National Laboratory for Physical Sciences at the Microscale, University of Science and Technology of China, Hefei, Anhui 230026, China}
\affiliation{CAS Center for Excellence in Nanoscience, University of Science and Technology of China, Hefei, Anhui 230026, China}

\author{Neil Qiang Su}
\affiliation{Department of Chemistry, Duke University, Durham, North Carolina 27708, USA}

\author{Weitao Yang} %\email{weitao.yang@duke.edu}
\affiliation{Department of Chemistry and Department of Physics, Duke University, Durham, North Carolina 27708, USA}
\affiliation{Key Laboratory of Theoretical Chemistry of Environment, School of Chemistry and
Environment, South China Normal University, Guangzhou 510006, China}

\date{\today}

\begin{abstract}

The delocalization error of popular density functional approximations (DFAs) leads to diversified problems in present-day density functional theory calculations.
For achieving a universal elimination of delocalization error, we develop a localized orbital scaling correction (LOSC) framework, which unifies our previously proposed global and local scaling approaches.
%[Phys. Rev. Lett. \textbf{107}, 026403 (2011), \emph{ibid.} \textbf{114}, 053001 (2015)].
%
The LOSC framework accurately characterizes the distributions of global and local fractional electrons, and is thus capable of correcting system energy, energy derivative and electron density in a self-consistent and size-consistent manner.
The LOSC--DFAs lead to systematically improved results, including the dissociation of cationic species, the band gaps of molecules and polymer chains, the energy and density changes upon electron addition and removal, and photoemission spectra.

\hfill \break
\end{abstract}

%\pacs{31.15.E-, 71.10.-w, 71.15.Mb}

\maketitle

%
% Introduction
%

\section{Introduction}
Despite the enormous success of density functional theory (DFT) \cite{Hohenberg64B864, Koh65A1133} in many fields of modern physics, its predictive power is impaired by the intrinsic errors associated with the approximations made for the exchange-correlation functional.
%It is thus critical to eliminate these errors universally to ensure the predictive power of DFT methods.
%
Delocalization error is one of the dominant errors of mainstream density functional approximations (DFAs). It is responsible for many failures of DFT calculations \cite{Mori-Sanchez08146401,Cohen08792,Cohen12289}.
Its physical origin is the violation of the Perdew--Parr--Levy--Balduz (PPLB) condition \cite{Perdew821691,Zhang00346,Yang005172,Perdew0740501}, which requires the total energy of a system as a function of electron number, $E(N)$, to be piecewise straight lines interpolating between integers. DFAs suffering from delocalization error
yield $E(N)$ underneath the piecewise linear segments for finite systems, and thus tend to give too low energy for delocalized electron distributions, and produce excessively delocalized electron distribution and sometimes qualitatively wrong density, as it falsely lowers the system energy \cite{Mori-Sanchez08146401,Cohen08792,Cohen12289}.
Moreover, the error in the total energy also transfers to the error in the energy derivatives with respect to the electron number, i.e., the chemical potentials. As a consequence, the frontier orbital energies as predicted by DFAs significantly deviate from the true ionization potentials or electron affinities.
This also applies to infinite systems, where delocalization error is indicated by the unphysical narrowing of the band gap.

\begin{figure}[t]
%\begin{centering}
%\includegraphics[scale=0.38]{error_illustr} %\par
\includegraphics[width=\linewidth]{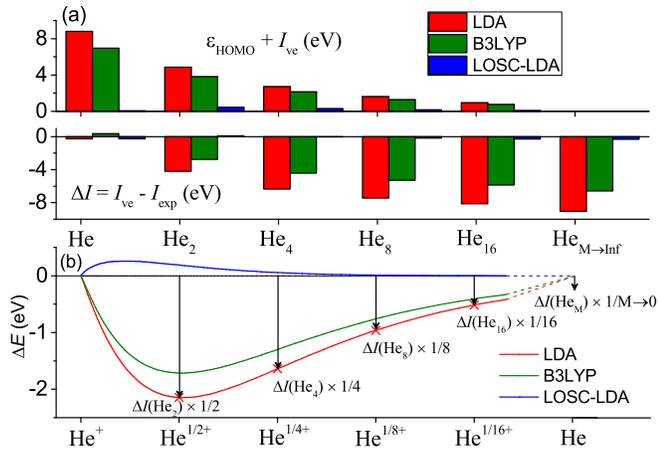}
%\end{centering}
\caption{ (a) Deviations between the calculated $\ep_{\rm HOMO}$ and $-I_{\rm ve}$ and between $I_{\rm ve}$ and $I_{\rm exp}$ for a series of He$_M$ clusters. In each cluster all the He atoms are chemically equivalent. The nearest neighboring atoms are separated by $10$\,\AA, and the $I_{\rm exp}$ of He$_M$ is well approximated by $I_{\rm exp}$ of a He atom.
%
%(b) Calculated total energy of a He atom versus fractional electron number $n$. The deviation of energy from the linearity condition,
%$\Delta E(n)=E(n)-nE(1)-(1-n)E(0)$, is shown, with $n=1$ representing the neutral He atom. See the text for more details. \label{fig1}}
(b) Calculated total energy deviation from the linearity condition of a fractionally charged He atom as a function of the fractional charge $\delta$. Here $\Delta E({\rm He}^{\delta+})=E({\rm He}^{\delta+})-\delta E({\rm He}^+)-(1-\delta)E({\rm He})$,
and the $\delta$ values have been scaled in the figure for a direct comparison with (a). \label{fig1}}
\end{figure}

The manifestation of delocalization error is size-dependent.
Figure~\ref{fig1}(a) explores the evolution of error in a series of loosely bound ${\rm He}_{M}$ clusters, where all the He atoms are chemically equivalent but separated by a large distance.
For a small $M$, for example $M=1$, the energy of the highest occupied molecular orbital ($\ep_{\rm HOMO}$) resulting from the paradigmatic local density approximation (LDA) significantly overestimates the calculated negative vertical ionization potential ($-I_{\rm ve}$) \cite{Cohen08115123},
leading to a large positive error bar (shown in red). In comparison, the deviation of $-I_{\rm ve}$ from the experimental value ($I_{\rm exp})$ is negligibly small.
As the cluster grows from the atomic limit ($M=1$) to the bulk limit ($M=\infty$), however, while $\ep_{\rm HOMO}$
gradually approaches $-I_{\rm ve}$, $I_{\rm ve}$ deviates increasingly from the experimental value ($I_{\rm exp}$), resulting in a large negative error bar.

The above deviations can be represented and understood from the fractional charge perspective \cite{Mori-Sanchez06201102, Mori-Sanchez08146401,Cohen08115123, Li17074107}.
Figure~\ref{fig1}(b) shows how the calculated $E(N)$ curve of a helium atom deviates from the PPLB condition.
%
%The LDA clearly results in a convex $E(N)$ curve, and $\ep_{\rm HOMO} = \frac{\partial E}{\partial N}|_{N\rightarrow 0^-}$ is the tangent slope at integer $N$.
%
The discrepancy $\Delta I= I_{\rm ve}-I_{\rm exp}$ for ${\rm He}_{M}$ corresponds quantitatively to the deviation of total energy from linearity, $\Delta E$, for a helium atom upon removal of $\frac{1}{M}$ electron. In particular, $\Delta E(\frac{1}{M})=\frac{1}{M}\Delta I({\rm He}_{M})$.
%
%Moreover, as $M\rightarrow \infty$, $\ep_{\rm HOMO} = \frac{\partial E}{\partial N}|_{N\rightarrow 0^-}= \lim_{M\rightarrow \infty} M [E({\rm He}^{1/M+})-E({\rm He})]=\lim_{M\rightarrow \infty} M \Delta I_M$ is the tangent slope at integer $N$.
%
Moreover, $\ep_{\rm HOMO}-I_{\rm ve}({\rm He}) = \frac{\partial E}{\partial N}|^{-}-I_{\rm ve}({\rm He})=-\frac{d \Delta E}{dx}\Big|_{x=0^+}$ is the tangent slope error at integer $N$ (note the positive $x$ direction in Figure~\ref{fig1}(b) is to the left),
which agrees with $-\Delta I({\rm He}_M)$ only in the infinite $M$ limit.
To see this, we calculate
\begin{align}
    -\frac{d \Delta E}{dx}\Big|_{x=0^+} &= -\lim_{M\rightarrow \infty} \frac{\Delta E(\frac{1}{M})-\Delta E(0)}{\frac{1}{M}} \nonumber \\
    &= -\lim_{M\rightarrow \infty} M \Delta E(\frac{1}{M}) \nonumber \\
    &= -\lim_{M\rightarrow \infty} \Delta I({\rm He}_M).
\end{align}
Therefore, the positive error bar in the orbital energy for $M=1$ exactly agrees in absolute value with the negative error bar in the total energy for $M=\infty$,
demonstrating that the delocalization error for finite (small) and bulk systems are similar but are manifested in two different ways \cite{Mori-Sanchez08146401}.

The situation is similar for many other DFAs, such as the popular hybrid B3LYP functional \cite{Becke935648,Lee88785}. As shown in \Fig{fig1}, inclusion of Hartree--Fock (HF) exchange only slightly reduces the energy deviations, because the delocalization error is compensated partly by the localization error (referring to $E(N)$ above the piecewise linear segment) associated with the HF exchange.

Here we make a comparison between the delocalization error with related concepts of self-interaction error (SIE) and many electron self-interaction error (MSIE).
SIE was the first concept to describe systematic error in DFAs and refers to the failure of compensation between the Hatree and exchange-correlation functional, or the error in the electron interaction energy, with DFAs in \emph{one-electron} systems. \cite{Perdew815048}
The SIE concept also applies to fractional electron systems with less than one electron. \cite{Zhang982604}
It had been presumed to be the key reason for some systematic failure in DFAs and various SIE--free functionals have been constructed. \cite{Zhang982604, Merkle929216, Perdew9716021, Bec05064101, Mori-Sanchez06201102, Vydrov0694108, Ruzsinszky06194112, Mori-Sanchez06091102, Scu08052513, Perdew815048, Vydrov048187, Vydrov05184107, Tao03146401, Tao0812509, Perdew08052513, Schmidt1414357, Liu12114104}
However, in 2006 this presumption was discovered untrue \cite{Mori-Sanchez06201102}: it was found that two well-developed functionals MCY2 \cite{Mori-Sanchez06091102} and B05 \cite{Becke0564101}, while both are SIE-free by construction and have improvements over hybrid functionals in describing chemical reaction barriers, still exhibit similar behavior on the remaining difficulties such as incorrect dissociation limit for molecular ions and overestimation of molecular polarizability for polymers, which was once associated with SIE.
This leads to the identification of a different source of the systematic errors of DFAs, namely, the essentially convex deviation from the exact linear condition of fractional number of electrons,
and the introduction of the concept of many-electron self-interaction error (MSIE). \cite{Mori-Sanchez06201102}
However, multiple definitions of MSIE exist: the term many-electron self-interaction error has also been defined in Ref \cite{Ruzsinszky06194112, Ruzsinszky07104102}, yet it was used to describe both convex and concave behaviors of fractional charges, which leads to totally different behaviors of errors in DFAs. \cite{Mori-Sanchez08146401}

The concepts of localization error (LE) and delocalization error (DE) were then introduced to capture the physical essence of the problem for systems with any number of electrons. \cite{Mori-Sanchez08146401}
Delocalization error, which exists in all commonly used DFAs including hybrids, describes the essentially convex deviation from the exact linear $E(N)$ curve for fractional charges and highlights the associated unphysical delocalization of electrons, or unphysically low energies for delocalized electrons. Localization error, which exists in Hartree-Fock (HF) approximation, describes the concave deviation from the exact linear $E(N)$ curve for fractional charges and highlights the associated unphysical localization of electrons.
Moreover, the concept of DE essentially incorporates the concept of SIE.
For one-electron systems, the SIE grows with increasing local fractional electron numbers and this is the reason for the failure in describing H$_{2}^{+}$ disscociation, as first pointed out in Ref \cite{Zhang982604}. This can also be characterized by DE.
The only missing component of SIE in the description of DE is the integer energy error for one electron systems that are compact as in molecular equilibrium structures. Nevertheless, for such systems, the SIE of commonly used DFAs is much smaller compared to fractional systems. \cite{Zhang982604} Neglecting the integer error, DE and SIE become identical for systems with one electron or less. In other word, functionals that are free of DE is essentially SIE-free, while SIE-free functionals can still suffer significantly from DE, just as other commonly used DFAs.

DE thus pinpoints the critical flaws in commonly used DFA and its reduction has been the driving force in many functional developments. As illustrated in Figure 1, most problems of DFAs relating to delocalization error can be quantitatively connected to the fractional E(N) curve of a single He atom. The corresponding localization error for HF is shown in the supplemental material. \cite{Supp}

%\cl{In the following discussions, we will focus on the perspective of LE/DE, in particular, the DE rather than LE since the former is the error that occurs in most semi-local and hybrid functionals.}
Enormous efforts have been devoted towards systematic removal of delocalization error. These include the development of global hybrid \cite{Becke935648, Adamo996158}, local hybrid \cite{Jaramillo031068, Arbuznikov07160}, doubly hybrid \cite{Grimme06034108, Zhang094963, Su149201}
and range-separated functionals \cite{Gill961005,Savin96,Leininger97151,Iikura013540,Baer1085,Stein10266802,Tsuneda10174101,Heyd038207,Yanai0451,
Baer05043002,Baer1085,Cohen07034101, Cohen07191109, Song07154105,Chai08084106}.
Like the B3LYP, the performance of these DFAs relies on the cancellation of errors, and is often system-dependent.
%
%There have also been efforts from the perspective of self-interaction error (SIE).
%
There are also attempts focusing on specific properties, such as the Koopmans-compliant functional \cite{Borghi14075135}, the generalized transition state method \cite{Anisimov05075125} and related approaches \cite{Ma1624924}, and others \cite{Chai13033002,Kraisler13126403,Gould13014103}.

To achieve a universal elimination of delocalization error, it is crucial to treat fractional electron distribution explicitly and properly. Following this idea, in 2011 we have developed a global scaling correction (GSC) approach by imposing the PPLB condition globally on any given system \cite{Zheng11026403, Zheng13174105, Zhang15154113}. The GSC largely restores the energy linearity condition at any fractional electron number, and thus reduces substantially the discrepancy between $\ep_{\rm HOMO}$ ($\ep_{\rm LUMO}$) and $-I_{\rm ve}$ ($-A_{\rm ve}$) for systems of all sizes.
Here, $\ep_{\rm LUMO}$ is the energy of the lowest unoccupied molecular orbital, and $A_{\rm ve}$ is the vertical electron affinity.
Retrieving PPLB condition for electron addition within the framework of DFT guarantees the improvement in the prediction of $-I_{\rm ve}$ using $\ep_{\rm HOMO}$, as well as $-A_{\rm ve}$ using $\ep_{\rm LUMO}$. \cite{Cohen08115123}
Despite the success of GSC, it does not offer any cure to the size-dependent discrepancy $\Delta I$, because it gives zero correction to energies at integers.
This implies that
(1) the parent functionals such as LDA is not \emph{size-consistent} for calculations of $I_{\rm ve}$ ($A_{\rm ve}$);
and (2) any functional correction that cannot affect
%any method correcting only the energies at global fractional electron numbers without affecting
the energies at integers is not \emph{size-consistent} either \cite{footnote0}.
To address this issue, in 2015 we have developed a local scaling correction (LSC) scheme that enforces the PPLB condition on local subsystems \cite{Li15053001, Zheng15581825}. The LSC is capable of correcting the total energy and electron density of integer--$N$ systems self-consistently, and thus leads to much improved description of dissociating molecules, transition-state species and charge-transfer systems.
However, the LSC has difficulty in capturing an infinitesimal amount of fractional electron, and so it cannot improve the prediction of $\ep_{\rm HOMO}$,  $\ep_{\rm LUMO}$ and thus fundamental gaps from the chemical potential differences \cite{Mori-Sanchez08146401, Cohen08792, Cohen12289}.

It would be ideal to incorporate the merits of describing global fractions (as in GSC) and local fractions (as in LSC) in one framework while avoiding their difficulties.
To this end, two major advancements are needed: (1) a unified scheme for characterizing global and local fractional electron distributions; and (2) an explicit and size-consistent correction
to system energies at both fractional and integer electron numbers.
%reduces delocalization error in the energies of systems with any number of electrons, integer and fractional.
%
In this paper, we develop a localized orbital scaling correction (LOSC) framework, which realizes these advancements.
To the best of our knowledge, this is the first density functional approximation that alleviates delocalization error in all aspects without involving system dependent parameters. In the rest of the paper, we will first revisit the forms of GSC and LSC, and then proceed to formulate the LOSC functional. Finally we will close with some concluding remarks.

\section{Motivation}
The basic idea of GSC is to add a correction functional to linearize each of the nonlinear components of the KS functional in the presence of fractional electron in the global system. The nonlinear components include the Hartree functional and the exchange-correlation functional,
with the former being a quadratic functional of the electron density $\rho$ and the latter having a more complicated scaling relation with $\rho$.
Nevertheless, by observation it has been found that the energy deviation from the linear $E(N)$ often displays a nearly parabolic shape, which suggests that one can approximately write the correction formula in terms of a quadratic expression of the fractional electron number as follows,
\begin{align}
  \Delta E^{\rm GSC} &= \frac{1}{2} \kappa \left( n_{\! f} - n_{\!f}^2 \right). \label{gsc-1}
\end{align}
Here $n_{\! f}=N-[N]$, and also equals the fractional electron occupation number on the KS frontier orbital, and $\kappa$ is a functional of the frontier orbital $\varphi_f(\br)$, which itself is a functional of the KS reduced density matrix $\rho_s$.
For LDA functional, in particular, $\kappa$ is approximated by the following form, \cite{Zheng11026403}
\begin{equation}
   \frac{1}{2}\kappa = \frac{1}{2}\! \iint\frac{\rho_{f}(\br)\rho_{f}(\br')}{|\br-\br'|}d\br d\br'- \frac{ C_{\rm x}}{3} \! \int[\rho_{f}(\br)]^{\frac{4}{3}}d\br,  \label{k-LDA}
\end{equation}
where $\rho_f(\br) = |\varphi_f(\br)|^2$ and $C_{\rm x} = \frac{3}{4}(\frac{6}{\pi})^{1/3}$.
One can verify that
$\frac{\partial \Delta E^{\rm GSC}}{\partial n_{\! f}}=\pm \frac{1}{2}\kappa$ at $n_{\! f} \rightarrow 0$ or $1$ gives a correction to
the tangent slope at integers, while $\frac{\partial^2 \Delta E^{\rm GSC}}{\partial n^2_{\! f}}=-\kappa$ modifies the curvature of $E(N)$ of the DFA.

The functional form of LSC can be viewed as a ``local'' version of \Eq{gsc-1},
\begin{align}
  \Delta E^{\rm LSC} & \approx  \frac{1}{2} \iint d\br d\br' \, \tilde{n}_{\! f}(\br) \left[1 - \tilde{n}_{\! f}(\br') \right]
  \tilde{\kappa}(\br,\br'),   \label{lsc-1}
\end{align}
with $\tilde{n}_{\! f}(\br)$ and $\tilde{\kappa}(\br,\br')$ being the local fractional electron occupation and local curvature, respectively.
These local variables are introduced as functionals of the KS reduced density matrix to capture the local fractional information. Note despite the fact that the original form of LSC as in Ref \cite{Li15053001} slightly differs from \Eq{lsc-1}, with some range separation and extra non-local attraction, \Eq{lsc-1} captures the main idea of LSC. Here for the simplicity of comparison, we stick with \Eq{lsc-1} as the approximate form of LSC.

%\emph{Unified characterization of global and local fractional electrons.}
%
Let us consider two typical examples: (1) H$_2^+$ at large internuclear distance $R$; (2) He$_M$ cluster as mentioned in Figure~\ref{fig1}. In (1), at large $R$, the one-electron H$_2^+$ molecule
yields a delocalized electron density over the two separated protons, with each subsystem density integrating to half an electron. The LDA severely underestimates the energy of the stretched H$_2^+$ because of delocalization error.
By design, GSC cannot capture the locally half electron information because it counts fractional electrons globally.
In contrast, LSC yields $\tilde{n}_{\! f}(\br) \simeq 0.5$ near each proton by imposing a spatial screening on the density matrix,
from which the local information is extracted, and then effectively performing the calculation of $n_f(
\br)-n_f(\br)^m$, with $n_f(\br)$ being the fractional component of the locally screened density matrix and $m$ a large integer (fixed to be 10). It is by the vanishing nature of high powers of any fractional number that enables $\tilde{n}_{\! f}(\br)$ to approximate $n_f(\br)$,
and thus capture the half electron information.
Yet, this only allows us to distinguish a fractional number from integer 0 or 1; it cannot distinguish two integers because $n_f(\br)-n_f(\br)^m=0$ identically for all such occupations.
Moreover, distinguishing a tiny fractional $\tilde{n}_{\! f}(\br)$ from zero is also numerically difficult.
This causes trouble in example (2). On the one hand, in the case of He$_M^+$, as $M\rightarrow \infty$ the local fraction becomes vanishingly small, which poses numerical challenge for LSC to capture the tiny fraction. On the other hand, in the case of neutral He$_M$ where local fraction is absent, LSC cannot capture the right ``local frontier orbital'' to compute the local curvature for the frontier orbital energy correction.
%As a remark, in the second case, GSC does not capture the ``local frontier orbital'' either. The global frontier orbital is captured, though.

The above analysis thus suggests that the key is to capture the ``local frontier orbital'' and its local occupation in order to solve all the problem together. Here we highlight that the two key words are ``local'' and ``frontier''. The extension from GSC to LSC captures ``local'' but overlooks ``frontier'', with the latter requiring energy information to enter in the local variable construction, rather than simply invoking the density matrix. This is a great challenge if our local extension is through the $\br$ space, since then we have to devise a ``local frontier energy'' function to achieve our purpose, which is difficult in both conceptual and practical manner.

\section{Formulation of LOSC}

We now pursue an alternative way to realize the local extension through the orbital space, by invoking localized orbitals (LOs).
Note that the KS density matrix is a sum over KS canonical orbital (CO) projections, $\rho_s = \sum_{m} n_m |\varphi_m\rangle \langle \varphi_m|$, and the occupation numbers $\{n_m\}$ are all integers (0 or 1) for integer systems.
The COs are not the only choice for unraveling $\rho_s$. Alternatively, we can exploit a localized representation of the density matrix as
\begin{equation}
    \rho_s = \sum_{ij} |\phi_i\rangle \langle \phi_i|\rho_s|\phi_j \rangle \langle \phi_j|=\sum_{ij}\lambda_{ij}|\phi_i\rangle\langle \phi_j|,
\end{equation}
%($\phi_i$'s being orthonormal), such that
%\begin{equation}
%    \rho(\br)= \rho_s(\br,\br)=\sum_{ij} \lambda_{ij} \phi_i(\br) \phi_j^\ast(\br), \label{LOSC-RDM}
%\end{equation}
where $\{\phi_i(\br)\}$ is chosen to be a set of orthonormal LOs, and $\lambda_{ij} = \langle \phi_i|\rho_s|\phi_j \rangle$ serves our purpose of being a local occupation matrix. In particular, one can show that the diagonal elements satisfy $0\leqslant \lambda_{ii}\leqslant 1$ and $\sum_i \lambda_{ii}=N$, so that each $\lambda_{ii}$ plays the role of an occupation number associated with $\phi_i$. Moreover, our desired local fractions arise naturally through the fractional components of $\{\lambda_{ii}\}$.
This is the motivation of LOSC. Now the remaining task is to construct the LOs from $\rho_s$ and to build a correction functional out of $\{\phi_i\}$ and $\lambda_{ij}$.

In the orbital space, the LOs should come from a unitary transformation of the COs through a localization procedure. In the traditional procedures, such as in the Boys \cite{Foster60300} and the Ruedenberg \cite{Edmiston63457} prescription, one minimizes a target function of only occupied COs, rendering the virtual COs irrelevant.
%\cl{In some recent works on periodic systems, the localized Wannier functions have been obtained from selected occupied subspaces as well as unoccupied subspaces. \cite{Mar121419}}
These approaches, however, cannot serve our purpose. In the H$_2^+$, for example, only one occupied CO exists, which leads to identical orbital with itself no matter what localization scheme is implemented. To obtain the desired LO, we have to incorporate the virtual COs into the localization scheme to form collective unitary rotations with the occupied orbitals. Moreover, we note that at large $R$, the HOMO and LUMO are near-degenerate, and energetically separated from other COs; the desired LOs should come from linear combinations (mixtures) of only HOMO and LUMO, and not from any other CO. Therefore, the localization procedure should be able to limit the mixing of these other orbitals with HOMO/LUMO.

This suggests that our desired localized orbitals should achieve \emph{a compromised localization both in the physical space and in the energy space}. In contrast, traditional localized orbitals keep localization only in the physical space by construction, while traditional canonical orbitals maintain localization only in the energy space, being energy eigenstates of an one-particle Hamiltonian. Here to better describe our desired localized orbitals, we introduce a new concept and call them \emph{orbitallets}, in analogy to wavelets, which achieve a compromised localization both in the physical space and in the momentum space. \cite{Daubechies92}

\begin{figure}[t]
\includegraphics[width=\columnwidth]{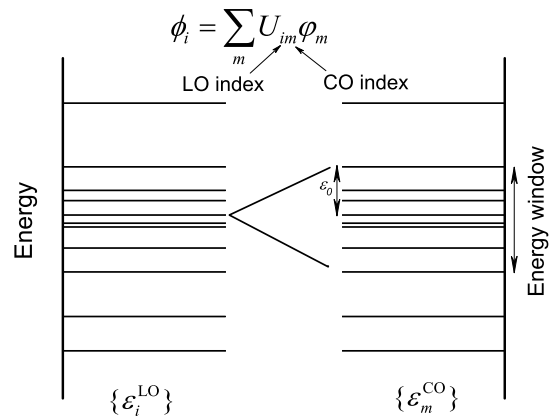}
\caption{
Schematic illustration of the relation between LOs and COs and the energy window.
Here each energy window is centered at $\ep_i^{\rm LO}$ with a fixed radius $\ep_0$. \label{fig2}}
%Here the LOSC is implemented in the post-SCF manner based on parent functional orbitals. The LOSC parameters are $R_0=2.7$\AA, $\eta=3.0$, $\epsilon_0=2.5$eV, $\tau=1.2378$, $\gamma=2.0$. Basis: 6-311++G(3df,3pd). Fitting basis: aug-cc-pVTZ.
\end{figure}

There are many ways to obtain localization in both the physical and energy spaces.
A simple way to achieve this is to modify the localization target function by adding a penalty function that enforces the localization of CO energies.
Here to have localization in energy space for our LOs (orbitallets), we have to define an energy spectrum for them. In this paper,
we define it to be the same as the CO spectrum.
One can invoke other definitions, but our definition seems to be the most natural choice.
Each LO with energy $\epsilon_i^{\rm LO}$ results from a mixture of the COs whose energies are within
a certain energy window of a fixed size centered at $\epsilon_i^{\rm LO}$, and the mixing coefficients come from a unitary matrix $U_{im} = \langle \phi_i | \varphi_m \rangle$.
See Figure~\ref{fig2} for a schematic illustration.
It is worth mentioning that localization involving both occupied and virtual orbitals have been used previously within a fixed energy window near the Fermi level for constructing maximally localized Wannier functions for systems with entangled energy bands. \cite{Souza01035109} Defined in a different way and for a different purpose, our localized orbitals, orbitallets, have dynamic energy windows opened at each $\epsilon_i^{\rm LO}$ and are designed to capture the local fractions in the system. 

In our present implementation of LOSC, %in particular,
$\{\phi_i(\br)\}$, the LOs, are generated by a restrained Boys localization procedure \cite{Foster60300, Supp}, which aims at minimizing the following spread function
\be
  F = \sum_i^\infty \Big[\langle \phi_i|\br^2|\phi_i\rangle - \langle \phi_i |\br|\phi_i\rangle ^2\Big] + \sum_{im}^\infty w_{im} |U_{im}|^2.  \label{boys-1}
\ee
%
%We align the LOs according to the energies of COs and introduce an energy window for the nearby COs around each energy level to mix into an LO.
Here the first term on the right hand side of \Eq{boys-1} resembles the Boys spread function, except that the sum rums over all the orbitals. In the second term,
a penalty function $w_{im}=w(|\epsilon_i^{\rm LO}-\epsilon_m^{\rm CO}|)$ is introduced to suppress mixing of orbitals beyond an energy radius $\epsilon_0$, in order to achieve energy localization.
As a remark, without the penalty term, minimization of $F$ with infinite number of orbitals (which span the entire Hilbert space of functions) will lead to a set of $\delta$ functions centered at different positions and $F=0$. Therefore, from pure mathematical perspective, the penalty function is also needed if the virtual orbitals participate in the localization. From the physical considerations, the penalty function $w(x)$ should satisfy the following requirements: (1) $w(0)=0$, which implies no penalty imposed against mixing between degenerate COs; (2) $w(x)$ is a monotonically increasing function; (3) $w(\infty)=\infty$, which forbids mixing between COs that are far apart in energy.
In this paper, we try to use the following simple function to achieve the above properties,
\begin{equation}
    w(x) = R_0^2 (\frac{x}{\epsilon_0})^\gamma [1-e^{-(\frac{x}{\epsilon_0})^\eta}],
\end{equation}
which involves four parameters. $R_0$ can be considered as related to the typical spread radius of small molecules, and is introduced here to factor out the length unit; $\epsilon_0$ is the energy window as mentioned above; and $\gamma$ and $\eta$ are the other parameters to adjust the shape of the function,
in particular, the asymptotic behavior as $x\rightarrow 0$ and $x\rightarrow \infty$.
Note that $\gamma$ has to be positive to satisfy condition (3). However, after some numerical experiment, we find that a positive $\gamma$ imposes insufficient penalty for $x<\epsilon_0$, leading to excessively artificial local fractions for compact molecules.
Therefore, we modify the function by dividing it into a piecewise function, with $\gamma=0$ when $x<\epsilon_0$ and $\gamma>0$ when $x\geqslant \epsilon_0$,
\begin{equation}
w(x)=
\left\{
\begin{array}{cl}
R_0^2 [1-e^{-(\frac{x}{\epsilon_0})^\eta}], & {\rm if}~x<\epsilon_0,\\
R_0^2 (\frac{x}{\epsilon_0})^\gamma [1-e^{-(\frac{x}{\epsilon_0})^\eta}],
 & {\rm if}~x\geqslant \epsilon_0.
\end{array}
\right. \label{w}
\end{equation}
The parameters are optimized for a balanced behavior on thermochemistry, reaction barrier heights and dissociation curves, resulting in $R_0=2.7$\AA, $\epsilon_0=2.5$eV, $\gamma=2.0$ and $\eta=3.0$, see the supplemental materials for more details. \cite{Supp}
The penalty function in \Eq{w} is continuous but not smooth at $x=\epsilon_0$. Yet this artificial feature has little impact on our results and can be easily removed by rounding it out or by a smooth interpolation between the two constructing functions. In the present work, since we put more emphasis on the general idea, we will not need to refine the details and make more sophisticated forms.
%$\lambda_{ij}=\langle \phi_i|\rho_s|\phi_j\rangle$ is an occupation matrix that conveys the information on fractional electron distribution.
%

\begin{figure}[t]
\includegraphics[width=\columnwidth]{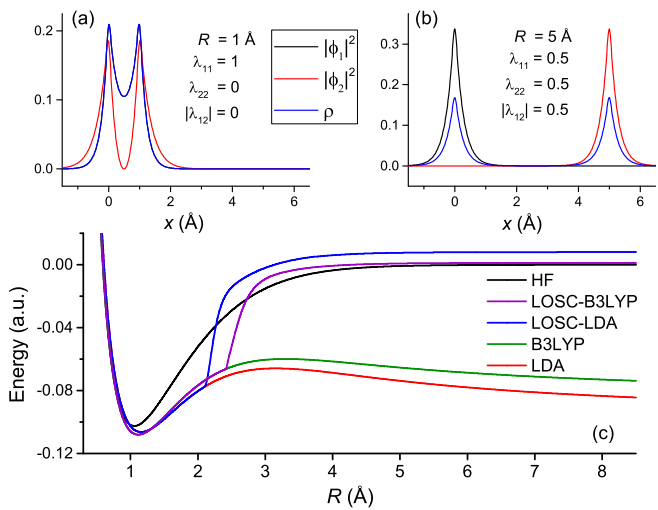}
\caption{
Distribution of LO densities along the bonding axis of H$_2^+$ at the internuclear distance of (a) $R=1$\,\AA~and (b) $R=5$\,\AA. The two protons locate at $x=0$ and $x=R$. The data are extracted from LDA calculations.
(c) Dissociation energy curve of H$_2^+$ calculated by various DFAs. The energy of a hydrogen atom is set to zero. \label{fig3}}
%Here the LOSC is implemented in the post-SCF manner based on parent functional orbitals. The LOSC parameters are $R_0=2.7$\AA, $\eta=3.0$, $\epsilon_0=2.5$eV, $\tau=1.2378$, $\gamma=2.0$. Basis: 6-311++G(3df,3pd). Fitting basis: aug-cc-pVTZ.
\end{figure}

In the minimal basis, the H$_2^+$ molecule has two KS orbitals. Figure~\ref{fig3}(a) depicts the LO densities of a compact H$_2^+$. Both LOs resemble the original COs, and the diagonal elements of the local occupation matrix remain integers, $\lambda_{11}=1$ and $\lambda_{22}=0$. This is because the HOMO and LUMO are far apart in energy so that their mixing is suppressed, rendering LOs of the similar character as the COs. This is reasonable since there is hardly any fractional electron distribution at a small internuclear distance $R$. In contrast, at a large $R$ the two nearly degenerate COs fully mix into two spatially separated LOs. The two LOs locate symmetrically at the two nuclei with $\lambda_{11} = \lambda_{22} = 0.5$, which reveals the fact that each proton carries half an electron (see Figure~\ref{fig3}(b)). In addition, the above behavior is almost independent of the basis set.

Here as a remark, for H$_2^+$ at a small $R$ or a large $R$, different choice of the localization target functions, for example Boys or Ruedenberg, does not make a difference in the optimized LOs, as long as the penalty function satisfies the three conditions. This, however, will make a difference in the intermediate $R$ or in a more complicated molecule. Yet this is a minor effect and beyond the scope of the present paper. Here we choose to modify Boys localization because it can be easily implemented and applied to systems of all sizes, and more importantly the Boys LOs can be replaced by Wannier functions \cite{Mar121419} for periodic solids.
%In contrast to all the conventional localization methods \cite{Foster60300, Mar121419} where only the occupied COs are considered, we here include both occupied and virtual COs in our localization to obtain $\{\phi_{i}(\br)\}$ to represent the density matrix in \Eq{LOSC-RDM}.
%The restrained Boys localization can be applied to systems of all sizes. For periodic solids the Boys LOs can be replaced by Wannier functions \cite{Mar121419}.
%The LOs and the associated $\bm\lambda$--matrix accurately characterize all the local fractional electron distributions in a system.
%Moreover, since $\sum_{i}\lambda_{ii}=N$ gives the exact total electron number, any global fractional electron is naturally captured by the LOSC framework.

\emph{Explicit functional form.} With the LOs (orbitallets) and $\bm\lambda$--matrix, an explicit form of LOSC is constructed, similar to \Eqs{gsc-1} and \eqref{lsc-1}:
\be
  \Delta E^{\rm LOSC} = \sum_{ij} \frac{1}{2}\kappa_{ij} \lambda_{ij}\left(\delta_{ij} - \lambda_{ij} \right) = \frac{1}{2} \text{tr} (\bm \kappa \bm \omega), \label{losc-1}
\ee
where $\omega_{ij} = \lambda_{ij}\left(\delta_{ij} - \lambda_{ij} \right)$.
Here the diagonal terms in the summation of \Eq{losc-1} is a natural extension of the energy correction formula of \Eq{gsc-1} from one global fraction to all the local fractions. Note that for $\lambda_{ii}=0$ or 1, their contributions to $\Delta E^{\rm LOSC}$ are zero and thus have no impact on the sum. The off-diagonal terms are introduced as non-local corrections to the unphysical interaction between the local fractions centered at different positions. They play similar role as the long-range attraction term in the LSC functional.
%
%\be
%  \Delta E^{\rm LOSC} = \sum_i \frac{1}{2}\kappa_{ii} \left(\lambda_{ii} - \lambda_{ii}^2 \right) - \sum_{i\neq j} \frac{1}{2} \kappa_{ij}   \lambda_{ij}^2. \label{losc-1}
%\ee
%
The curvature matrix elements $\{\kappa_{ij}\}$ for LDA and the generalized gradient approximations (GGAs) are calculated via
\begin{equation}
   \frac{1}{2}\kappa_{ij} = \frac{1}{2}\! \iint\frac{\rho_{i}(\br)\rho_{j}(\br')}{|\br-\br'|}d\br d\br'- \frac{\tau C_{\rm x}}{3} \! \int[\rho_{i}(\br)]^{\frac{2}{3}}[\rho_{j}(\br)]^{\frac{2}{3}}d\br,  \label{k-mat-1}
\end{equation}
where $\rho_i(\br)=|\phi_i(\br)|^2$ is the $i$th LO density.
This is a straightforward extension of \Eq{k-LDA} in a symmetric manner with respect to $i$ and $j$, but with a parameter $\tau$. In \Eq{k-LDA}, $\tau$ is set to 1, which is good for the orbital corrections. In particular, in the case of HOMO energy correction of a hydrogen atom, with the exchange only LDA functional one can show that \Eq{k-LDA} exactly compensates the wrong slope under the frozen orbital assumption. \cite{Supp} However, $\tau=1$ does not retrieve the right amount of correction for H$^{1/2+}$. This is because the LDA exchange functional is not quadratic, so that the energy deviation from linearity in the $E$ vs $N$ curve of (exchange only) LDA is not strictly quadratic either, although it can be approximately treated as a parabola. As a consequence, under this parabolic assumption, one cannot simultaneously retrieve the right slope at integer and the energy at half integer. In order to do that, higher order corrections have to be introduced. \cite{Zhang15154113} In the present paper, we are biased towards the half integer energy within the parabolic correction under the frozen orbital analysis, and
assign $\tau$ a nonempirical value of $\tau = 6(1-2^{-1/3})\approx 1.2378$ \cite{Supp}.
In the frozen-orbital approximation, \Eq{losc-1} leads to the following correction to the CO energies \cite{Supp}:
\begin{equation}
   \Delta \epsilon_m = \sum_i \kappa_{ii} \left( \frac{1}{2}-\lambda_{ii} \right) |U_{im}|^2
   - \sum_{i\neq j} \kappa_{ij} \lambda_{ij} U_{im} U_{jm}^\ast.  \label{delta-ep-1}
\end{equation}

As shown in \Fig{fig3}(c), the dissociation energy curve of H$_2^+$ is greatly improved for $R > 2$\,\AA~by using \Eq{losc-1}; while for more compact geometries ($R < 2$\,\AA) the energy correction is almost zero. The latter is due to the specific form of the penalty function adopted in the restrained Boys localization of \Eq{boys-1} -- it is designed to preserve the good accuracy of parent DFAs on thermochemistry for small- and medium-sized molecules.
In particular, at a small $R$, the localization plays trivial role in the sense that the unitary transformation $\bm U$ is almost an identity matrix. This suggests that the minimizer of the restrained Boys target function is achieved almost at its boundary due to the small overall spread of COs and large penalty against their mixing. As $R$ increases, the CO spread grows while the penalty term is alleviated. When $R$ reaches a critical value (which is mainly dependent on the $R_0$ parameter; smaller $R_0$ will reduce the critical $R$), mixing between COs becomes favorable so that $\bm U$ becomes different from identity. This causes a kink at the critical $R$, an artifact due to the choice of our present localization scheme, and shall be addressed through a better construction scheme in future work.
In addition, we note that for large $R$, the LOSC-LDA energy is slightly above zero, which suggests that H$^{1/2+}$ is over-corrected. This is because (1) the LDA correlation energy correction has been neglected- we introduce a non-empirical $\tau$ parameter only to account for the LDA exchange curvature; (2) the unrelaxed LDA COs in the post-SCF implementation leads to some overcorrection, which is another effect in the design of the $\tau$ parameter. Self-consistent implementation could only modestly relax the energy, without achieving the zero energy limit; in order to reach the zero energy limit, one has to go beyond the frozen orbital assumption in the design of curvature matrix and total energy correction formula.

One can also apply LOSC to other more accurate parent functionals. In this paper, we have achieved this
by designing flexible forms of $\kappa_{ij}$ \cite{Supp} on the basis of the curvature formula of LDA for many other types of DFAs , including the GGAs, the hybrids such as the B3LYP, the range-separated functionals such as the CAM-B3LYP \cite{Yanai0451}, etc. These DFAs suffer from the delocalization error to different extents, while the LOSC gives similar corrected results; see for instance the H$_2^+$ dissociation curves calculated by LOSC--LDA and LOSC--B3LYP in \Fig{fig3}(c). Moreover, the fact that LOSC-B3LYP energy at large $R$ is almost perfect suggests that LOSC applied to a better parent functional leads to better results.
%
%Note that the LOSC--LDA slightly overcorrects the energy in the dissociation limit ($R \rightarrow \infty$).

In a related effort, Anisimov and Kozhevnikov have developed a generalized transition state (GTS) method to improve the LDA calculation for band gaps of solids \cite{Anisimov05075125}. Their suggested energy correction amounts to $\Delta E^{\rm GTS} = \sum_{i} \frac{1}{2}\kappa_{ii} (\lambda_{ii}-\lambda_{ii}^2)$, where each $\kappa_{ii}$ is determined by a separate constrained LDA calculation.
A similar scheme was recently constructed by Ma and Wang \cite{Ma1624924}.
In these works the LOs come from mixing of only occupied or virtual COs in the localization and their energy corrections, thus do not change the total energies for physical systems with integer number of electrons; hence these energy functionals are not \emph{size consistent} \cite{footnote0} and can only correct orbital energies.
These methods have only been implemented as post-DFT corrections rather than in a self-consistent manner, thus they rely on the qualitatively correct DFT densities to produce reasonable corrections.
In contrast, the LOSC framework uses mixing of the occupied and virtual COs in the localization and offers an explicit form of \Eq{k-mat-1} for computing the $\bm\kappa$--matrix.
%
%The LOSC framework uses both the occupied and virtual COs in the localization and offers an explicit form of \Eq{k-mat-1} for computing the $\bm\kappa$--matrix.
%
%if only occupied COs were included in the localization and thus do not change the total energies for physical systems with integer number of electrons; thus these energy functionals are not \emph{size consistent} \cite{footnote0}. %In contrast,
%
It changes the DFA energies both at integer and fractional electron numbers.
Moreover, \Eq{losc-1} involves off-diagonal $\kappa_{ij}$ and $\lambda_{ij}$ that are crucial because they dispel the unwanted interactions between LO pairs. In the case of H$_2^+$ dissociation, it is only with these off-diagonal terms that the correct asymptotic behavior as $R \rightarrow \infty$ can be retrieved.
Importantly, LOSC is a functional of the non-interacting density matrix and can be implemented self-consistently within the generalized Kohn-Sham approach.

\emph{Self-consistent corrections to energy and electron density.}  For practical calculations we have devised a self-consistent field (SCF) procedure, with which the LOSC approach improves $E(N)$ and $\rho(\br)$ simultaneously.
The SCF procedure consists of a series of steps as follows,
\begin{equation}
    \bm\rho_s^{\rm in} \overset{(\rm I)}{\longrightarrow} \bm h_p \overset{(\rm {II})}{\longrightarrow} \{ \varphi_m \} \overset{(\rm {III})}{\longrightarrow} \{\phi_i\}
    \overset{(\rm{IV})}{\longrightarrow} \Delta \bm h \overset{(\rm{V})}{\longrightarrow} \bm\rho_s^{\rm{out}}. \label{scf-1}
\end{equation}
In step (I) a projected KS (or generalized KS, GKS) Hamiltonian is constructed from the initial density matrix as
$\bm h_p = \bm\rho_s \bm h_0 \bm \rho_s + (\bm I- \bm\rho_s)\bm h_0 (\bm I- \bm\rho_s)$, with
$\bm h_0 = \frac{\delta E^{\rm DFA}}{\delta \bm \rho_s}$
being the KS/GKS Hamiltonian of the parent DFA.
Here, the projections on $\bm h_0$ using $\bm\rho_s$ and $\bm I- \bm\rho_s$ avoid overcorrecting the energies of compact molecules
by rendering $\lambda_{ii}$ close to integer 0 or 1.
\cite{Supp}.
In step (II) $\bm h_p$ is diagonalized to generate the auxiliary COs.
In step (III) the restrained Boys localization is carried out to obtain the LOs. In step (IV) the LOSC contribution to GKS Hamiltonian matrix is computed via $\Delta\bm h = \frac{\delta \Delta E^{\rm LOSC}}{\delta \bm \rho_s}$.
Finally in step (V) $\bm \rho_s$ is updated by minimizing the total energy with the aid of approximate gradient $\bm h_0 + \Delta\bm h$ \cite{Supp}.
Steps (I)--(V) are iterated until the initial and final density matrices are equal.
It is worth pointing out that although the LOSC involves orbitals, it remains a functional of the density matrix and its SCF implementation can be carried out within the GKS scheme (or the Hartree-Fock-Kohn-Sham scheme) \cite{Parr89}, because the Hamiltonian $\bm h_p$ and the COs and LOs are all determined by $\bm\rho_s$.
Thus, $\ep_{\rm HOMO}$/$\ep_{\rm LUMO}$ of the LOSC hamiltonian $\bm{h}_{0}+\Delta\bm{h}$ is the chemical potential for electron removal/addition and the HOMO--LUMO gap is the derivative gap, which are theoretical predictions of the fundamental gap in both finite molecules and bulk \cite{Cohen08115123, Mori-Sanchez08146401, Cohen12289}.

\begin{figure*}[t!]
%\begin{centering}
%\includegraphics[scale=0.32]{Gap} \par
\includegraphics[width=\textwidth]{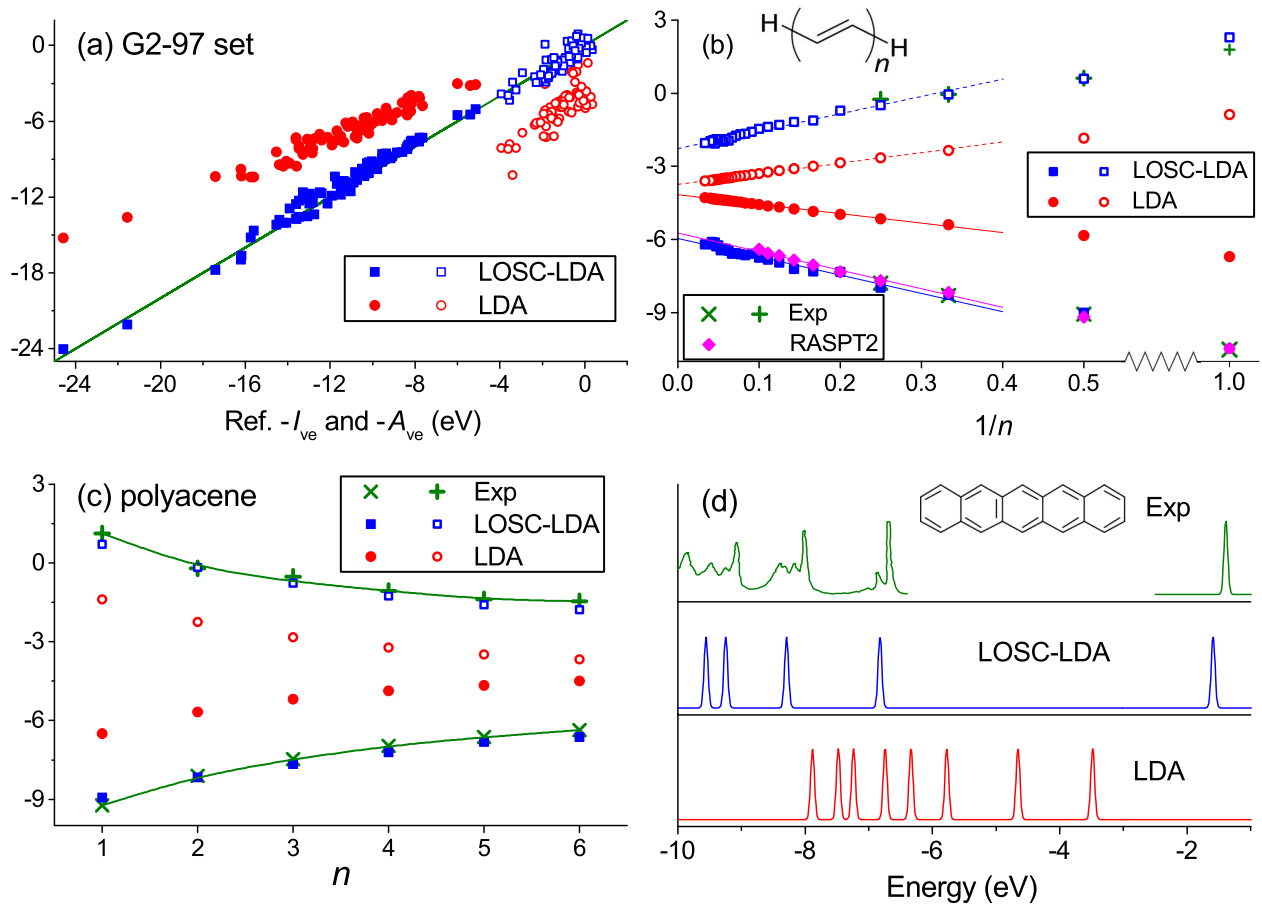}
%\end{centering}
\caption{
%(a) Calculated $\epsilon_{\rm HOMO}$ versus $-I_{\rm ve}$  ($\epsilon_{\rm LUMO}$ versus $-A_{\rm ve}$) for 70 (47) molecules of the G2--97 set \cite{Curtiss971063} and for atoms H--Ar, as represented by the solid (hollow) symbols \cite{footnote1}.
%The mean absolute deviation is 4.37 (3.57)\,eV with the LDA, and 0.47 (0.43)\,eV with the LOSC--LDA.
The y axes in (a)-(c) stand for $\epsilon_{\rm HOMO}$ ($\epsilon_{\rm LUMO}$) in unit of eV, and in (d) stands for density of states (DOS) in atomic unit. In (a)-(c), $\epsilon_{\rm HOMO}$ ($\epsilon_{\rm LUMO}$) are represented by the solid (hollow) symbols.
The plot details are as follows.
(a) Calculated $\epsilon_{\rm HOMO}$ ($\epsilon_{\rm LUMO}$) of LDA and LOSC-LDA for 82 (61) atoms and molecules in a modified G2--97 set \cite{Curtiss971063,footnote1}, in comparison with the reference $-I_{\rm ve}$ ($-A_{\rm ve}$) \cite{footnote2, Pearson88734}.
The mean absolute error (MAE) is 4.43 (3.76)\,eV with the LDA, and 0.50 (0.50)\,eV with the LOSC--LDA.
(b) Calculated $\epsilon_{\rm HOMO}$ versus $-I_{\rm exp}$ ($\epsilon_{\rm LUMO}$ versus $-A_{\rm exp}$) for trans-polyacetylene oligomers. %H(C$_2$H$_2$)$_n$H.
The reference experimental values and the second order restrictive active space perturbation theory (RASPT2) calculated values are obtained from Refs.~\cite{Shahi0910964, Puiatti081394, Jordan87557, Allan841776}.
(c) Calculated $\epsilon_{\rm HOMO}$ versus $-I_{\rm exp}$ ($\epsilon_{\rm LUMO}$ versus $-A_{\rm exp}$) for polyacene oligomers.
The green lines in (a) and (c) are guide to the eyes. The reference experimental data are given by
Refs.~\cite{NIST, Burrow879, Hajgato08084308}.
%Refs.~[\onlinecite{Curtiss9842,NIST,Stu703458,Bou701279,Poo755179,Son96,Mad04}].
%
(d) DOS spectrum of pentacene in comparison with experimental photoemission spectrum \cite{Boschi72116}.
All peaks in the calculated DOS and the peak at -1.4\,eV in the experimental spectrum are broadened by Gaussian functions for clarity.
\label{fig4}}
%The LOSC parameters are $R_0=2.7$\AA, $\eta=3.0$, $\epsilon_0=2.5$eV, $\gamma=2.0$ and $\tau=1.2378$. \label{fig4}}
\end{figure*}

%\emph{Extensive applicability and enhanced efficiency.} The LOSC approach is applicable to many types of DFAs, including the LDA, the GGAs, the hybrids such as the B3LYP, the range-separated functionals such as the CAM-B3LYP \cite{Yanai0451}, etc. This is achieved by designing flexible forms of $\kappa_{ij}$ \cite{Supp}. These DFAs suffer from the delocalization error to different extents, while the LOSC gives similar corrected results; see for instance the H$_2^+$ dissociation curves calculated by LOSC--LDA and LOSC--B3LYP in \Fig{fig2}(c).

The implementation of LOSC is very efficient, since the computation of pertinent quantities such as $\bm\lambda$ and $\bm \kappa$ are straightforward. The restrained Boys localization can be conducted efficiently using the Jacobi sweep approach \cite{Edmiston63457, Barr75537}.
%Moreover, the orbitals far apart from the Fermi level have rather little contribution to the $\Delta E^{\rm LOSC}$ and thus need not be incorporated into the LOSC framework.
%\cl{Furthermore, the $\Delta \bm h$ in (\ref{scf-1}) has been approximated for numerical and efficiency considerations. \cite{Supp}}
Consequently, the extra computational cost due to the LOSC procedure usually amounts to a small portion of the overall cost.

\section{Results and Conclusion}

In the following, we demonstrate that the LOSC approach generally alleviates the delocalization error associated with the mainstream DFAs,
and thus cures many related problems in practical DFT calculations.

To start with, the size-dependent deviations between the calculated $\ep_{\rm HOMO}$ and $-I_{\rm ve}$
and between $I_{\rm ve}$ and $I_{\rm exp}$ %for the He$_M$ clusters
are mostly eliminated by applying the LOSC, suggesting that LOSC achieves size-consistency; see \Fig{fig1}(a).
As indicated in \Fig{fig1}(b), the LOSC largely straightens the $E(N)$ curve between integers.
Furthermore, the straightening of $E(N)$ curve is achieved not only for electron removal (related to HOMO prediction), but also for electron addition (related to LUMO prediction); not only for compact systems, but also for dissociating molecules such as He$_2^+$. \cite{Supp} In the latter case, when the parent DFA predicts wrong integer energy, merely straightening the $E(N)$ curve does not cure the problem. In such case, the LOSC not only straightens the curve, but also shifts the integer energy so as to point to the right slope, see for instance the $E(N)$ curve connecting He$_2$ and He$_2^+$ at $R=5$\AA. \cite{Supp}

Besides the H$_2^+$ shown in \Fig{fig2}(c), the LOSC also systematically improves the dissociation behavior of many other molecular cation species, such as the He$_2^+$, the water dimer cation, and the benzene dimer cation \cite{Supp}.
This suggests that the LOSC does not deteriorate with system size.

The LOSC systematically improves the prediction of $\ep_{\rm HOMO}$, $\ep_{\rm LUMO}$ and thus the fundamental gaps for systems of all sizes, ranging from atoms and molecules to polymers such as polyacenes and trans-polyacetylenes; see \Fig{fig4}.
%
%In the bulk limit the $E(N)$ curve becomes perfectly straight, and it is the wrong slope that leads to the inaccurate gaps.
%Consequently, the GSC has no effect on gaps infinite systems such as polymers \cite{Zheng11026403}.
In contrast to GSC which has no effect (and LSC which has numerical difficulty) on band gaps of bulk, \cite{Zheng11026403} the LOSC gives a promising improvement as it has a nonzero correction on polymers in the extrapolated infinite
chain length limit.
%
%For calculation of solids we adopt a modified wannier90 program \cite{Mos08685} to generate the Wannier functions via the restrained localization \cite{Supp}.
%
%As shown in \Fig{fig4}(b), the LOSC--LDA correctly predicts zero gaps for metals and retrieves large gaps for insulators, while the gaps of semiconductors are somewhat overcorrected.
%
%Such overcorrection is likely due to the incompatibility between the molecular penalty matrix $\bm w$ and periodic boundary condition, and may be avoided by designing a more reasonable form of $\bm w$--matrix.

In addition to $\ep_{\rm HOMO}$ and $\ep_{\rm LUMO}$, the LOSC also corrects the energies of other KS orbitals via \Eq{delta-ep-1}. As depicted in \Fig{fig4}(d), the orbital energies of pentacene predicted by the LOSC--LDA agree accurately with the peak positions in experimental photoemission spectrum.
This feature will be further studied in our subsequent papers.

\begin{figure}[t]
%\begin{centering}
\includegraphics[width=\columnwidth]{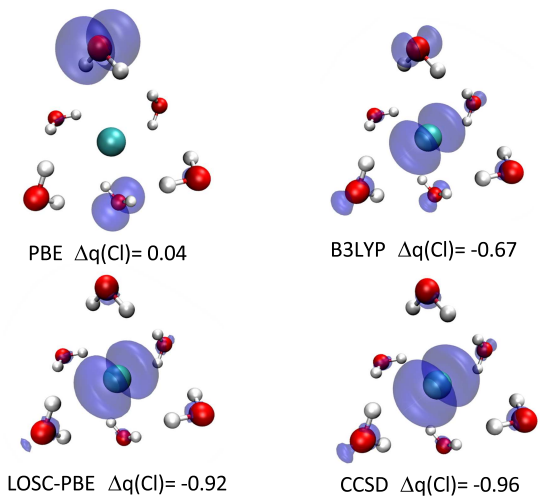}
%\end{centering}
\caption{Calculated charge density difference between a solvated Cl$^-$ and a solvated Cl atom with different methods.
The structure model consists of a Cl atom surrounded by six water molecules.
The isosurface of $0.003\,$\AA$^{-3}$ is shaded in blue, and $\Delta q$ is the change in Mulliken atomic charge
on Cl upon electron addition.
The coupled cluster method with single and double excitation (CCSD) \cite{Cizek69,Purvis82,Pople87} is taken as the reference.
%Here the six water molecules are placed equivalently around the Cl atom. ?? equivalent?
\label{fig5}}
\end{figure}

Finally, the LOSC also corrects the wrong electron density caused by the delocalization error. A typical example is a solvated Cl anion \cite{Cohen08792,Cohen12289}.
As shown in \Fig{fig5}, the PBE functional \cite{Perdew963865} erroneously predicts that the excess electron delocalizes over several water molecules, while the LOSC--PBE yields the correct distribution that the excess electron mostly localizes on the Cl atom.

The examples shown demonstrate that the LOSC approach remedies a wide range of problems caused by the delocalization error of DFAs.
The LOSC functional is very different from conventional density functional constructions, which are explicit analytical functionals of the density, the density gradients or the KS reduced density matrix. The LOSC framework utilizes many other information, such as the localized orbitals , local occupation matrix and parent DFA reference spectrum, which themselves are implicit functionals of $\rho_s$. This opens up a lot more possibilities in the exploration of the exact functional within the functional space.
As a first effort, the LOSC functional presented in this paper addresses the size-consistency problem and exhibits a systematic
elimination of delocalization error in all aspects. With a more sophisticated localization procedure and a more complete form of $\Delta E^{\rm LOSC}$ addressing local fractional spins, it is possible to eliminate other intrinsic errors of the mainstream functionals, and systematically improve the density functional approximation to a greater extent.

%Remaining challenges concern the treatment of fractional electron distributions in compact molecules and in periodic solids.
%
%These would require designing of a more sophisticated localization procedure for LOs, a more accurate form of $\Delta E^{\rm LOSC}$, and a more efficient SCF scheme.

%
% acknowledgment
%

Support from the National Science Foundation (CHE-1362927) (CL), the Ministry of Science and Technology of China (Grants No.\,2016YFA0400900 and No.\,2016YFA0200600) (XZ),
the National Natural Science Foundation of China (Grants No.\,21573202 and No.\,21233007) (XZ),
the Strategic Priority Research Program (B) of the Chinese Academy of Sciences (XDB01020000) (XZ),
the Fundamental Research Funds for Chinese Central Universities (Grant No.\,2340000074) (XZ),
and the Center for Computational Design of Functional Layered Materials (Award DE-SC0012575),
an Energy Frontier Research Center funded by the US Department of Energy, Office of Science, Basic Energy Sciences.
(NQS, WY) is appreciated.
Discussions with Dadi Zhang, Prof. Jianfeng Lu and Dr. Tomasz Janowski were helpful.

%\bibliographystyle{nsr}
%\bibliographystyle{aip}

%\bibliography{database_total}

\end{document}

\draft

Supporting Information of
%
%\vspace{0.2cm}
\vspace{0.1cm}

\begin{center}
{\large \textbf{Localized Orbital Scaling Correction for Systematic Elimination of Delocalization Error in Density Functional Approximations}}
\vspace{0.2cm}

Chen Li$^1$, Xiao Zheng$^2$$^,$$^3$, Neil Qiang Su$^1$
 and Weitao Yang$^4$$^,$$^5$

{\small
\emph{$^1$Department of Chemistry, Duke University, Durham, North Carolina 27708, USA
\\}
\emph{$^2$Hefei National Laboratory for Physical Sciences at the
Microscale, University of Science and Technology of China, Hefei, Anhui
230026, China\\}
\emph{$^3$Synergetic Innovation Center of Quantum Information and
Quantum Physics, University of Science and Technology of China, Hefei, Anhui
230026, China\\}
\emph{$^4$Department of Chemistry and Department of Physics, Duke University, Durham, North
Carolina 27708, USA\\}
\emph{$^5$Key Laboratory of Theoretical Chemistry of Environment, School of Chemistry and
Environment, South China Normal University, Guangzhou 510006, China}
}

\vspace{0.5cm}
%\today

(Dated: \today)

\end{center}

\linespread{1.0}

%\vspace{0.2cm}

\tableofcontents

\clearpage

%%%%%%%%%%%%%%%%%%%
\renewcommand{\theequation}{S\arabic{equation}}
\renewcommand{\thetable}{S\arabic{table}}
\renewcommand{\thefigure}{S\arabic{figure}}
%%%%%%%%%%%%%%%%%%%%%

\section{Elaboration on size-consistency requirement in terms of ionization energy and Electron affinity}

Given two non-interacting system $A$ and $B$, size consistency requires the energy of the super system of $A\cdots B$, at the infinite separation limit, to be equal to the sum of energies of isolated system $A$ and $B$, i.e.,
\begin{equation}
    E(A\cdots B) = E(A) + E(B). \label{size-con-1}
\end{equation}
This requirement is often applied to neutral systems. Here we extend the size-consistency condition to general subsystems with arbitrary number of electrons. Let $E(Z; N_Z)$ ($Z=A$ or $B$) be the ground state energy of an isolated system $Z$ with $N_Z$ (integer) electrons. Then the size consistency condition for the super system $A\cdots B$ with $N=N_A+N_B$ electrons is that
\begin{equation}
    E(A\cdots B; N) = \min_{N_A+N_B=N} E(A; N_A) + E(B; N_B). \label{size-con-2}
\end{equation}
As a special case, for any neutral system $Z$ with $N_Z$ electrons, consider a super system of $Z \cdots Z \cdots Z$ with $M$ subsystems, let us denote as $Z_M$. Then by condition \eqref{size-con-1}, we have
\begin{equation}
    E(Z_M; N) = M E(Z; N_Z). \label{neutral}
\end{equation}
Here $N= M N_Z$. Moreover, with a vertical ionization of $Z_M$ and by applying condition \eqref{size-con-2} $(M-1)$ times, we have
\begin{equation}
    E(Z_M; N-1) =  \min_{\sum N_i=N-1 } \sum_{i=1}^M E(Z; N_i)=(M-1)E(Z; N_Z) + E(Z; N_Z-1). \label{cation}
\end{equation}
For the second equality, we have used the assumption that (i) the integer energy is convex, i.e., $E(Z; M-1)+E(Z; M+1) > 2E(Z; M)$ for any integer $M$; and (ii) the energy of a fractional electron system satisfies the PPLB condition, i.e., $E(Z, N_i) = (1-\delta) E(Z, [N_i]) + \delta E(Z, [N_i]+1)$, with $[N_i]$ being the largest integer no greater than $N_i$, and $\delta = N_i-[N_i]$.

By subtracting \Eq{neutral} from \Eq{cation}, we have
\begin{align}
    I_{\rm ve}(Z_M) &= E(Z_M; N-1)-E(Z_M; N) \nonumber \\
    &= E(Z; N_Z-1)- E(Z; N_Z) \nonumber \\
    &= I_{\rm ve}(Z), \label{size-I}
\end{align}
which means that the vertical ionization potential of the super system $Z_M$ is constant as $M$ varies (an intensive quantity). Similar argument applies to the electron affinity as well, i.e.,
\begin{equation}
    A_{\rm ve}(Z_M) = A_{\rm ve}(Z), \label{size-A}
\end{equation}
Therefore, Eqs.~\eqref{size-I}--\eqref{size-A} are the size-consistency conditions expressed in terms of intensive quantities like $I$ and $A$.

Following the discussion in Ref.~\onlinecite{Mori-Sanchez08146401}, we can show that any DFA with delocalization error is not size-consistent, meaning Eqs. \eqref{size-I}--\eqref{size-A} are not satisfied. For $M=1$, common DFAs give a good vertical ionization energy $I_{\rm ve}(Z)$, but $-\epsilon_{\rm{HOMO}}$ underestimates $I_{\rm ve}(Z)$. For finite $M$, \Eq{neutral} holds for DFA on neutral systems, and the total energy is size-consistent. But with the removal of one electron, the hole generated is delocalized and each subsystem is a fractional charge (hole) system with the fractional charge (hole) integrated to 1-$\frac{1}{M}$ ($\frac{1}{M}$). Thus $E(Z_M; N-1) = M[E(Z; N_Z-\frac{1}{M})]$. As $M \rightarrow \infty$,
\begin{equation}
    E(Z_M; N-1) \rightarrow M\Big[E(Z; N_Z) -  \frac{1}{M} \frac{\partial E(Z; N_Z)}{\partial N_Z}\Big|^- \Big] = E(Z_M; N) - \epsilon_{\rm HOMO}(Z; N_Z),
\end{equation}
where $\epsilon_{\rm HOMO}(Z; N_Z)$ is the HOMO energy of an isolated subsystem. It follows that the vertical ionization energy becomes
\begin{equation}
    I_{\rm ve}(Z_M) = E(Z_M; N-1) - E(Z_M; N) \rightarrow - \epsilon_{\rm HOMO}(Z; N_Z),
\end{equation}
which underestimates $I_{\rm ve}(Z)$. Thus $I_{\rm ve}(Z_M)$ violates the size-consistency condition.

Upon improving a DFA with delocalization error, if the correction to total energy is finite for fractional number of electrons but zero for integer number of electrons (as in the case of GSC \cite{Zheng11026403}, GTS method \cite{Anisimov05075125} and a similar approach recently proposed by Ma and Wang \cite{Ma1624924}),
then such corrected functionals do not change the preceding argument for parent DFA and its violation of size consistency conditions
Eqs.~\eqref{size-I}--\eqref{size-A} remains unchanged.
However, with post SCF corrections to a DFA, such as the GTS and the work by Ma and Wang,
while the energy at the integer numbers remain unchanged, the correction to the chemical potential $\frac{\partial E(Z_M; N)}{\partial N}\Big|^-$ is finite as $M\rightarrow \infty$, and can lead to reasonable correction to %the calculated derivative gaps.
the calculations of $I$ and $A$ from chemical potentials, but not from total energy finite difference calculations.

The above analysis justifies the observations in Figure 1 of the main text. In particular, LDA and B3LYP (and all other DFAs that suffer from delocalization error) are not size-consistent when calculating $I_{\rm ve}$ and $A_{\rm ve}$, although they are size-consistent in terms of total energy of a neutral system. In contrast, the LOSC achieves size-consistency for both total energy and $I_{\rm ve}$ ($A_{\rm ve}$). Here as a remark, the LSC largely restores the size-consistency by maintaining $I_{\rm ve}(Z_M) = I_{\rm ve}(Z)$ for finite $M$ ($M \leqslant 10$ by design), however, the size-consistency issue occurs for large $M$ because of the numerical difficulty in capturing tiny fractions.

\section{Details about restrained Boys localization}

\begin{figure}[t]
\begin{centering}
\includegraphics[scale=0.4]{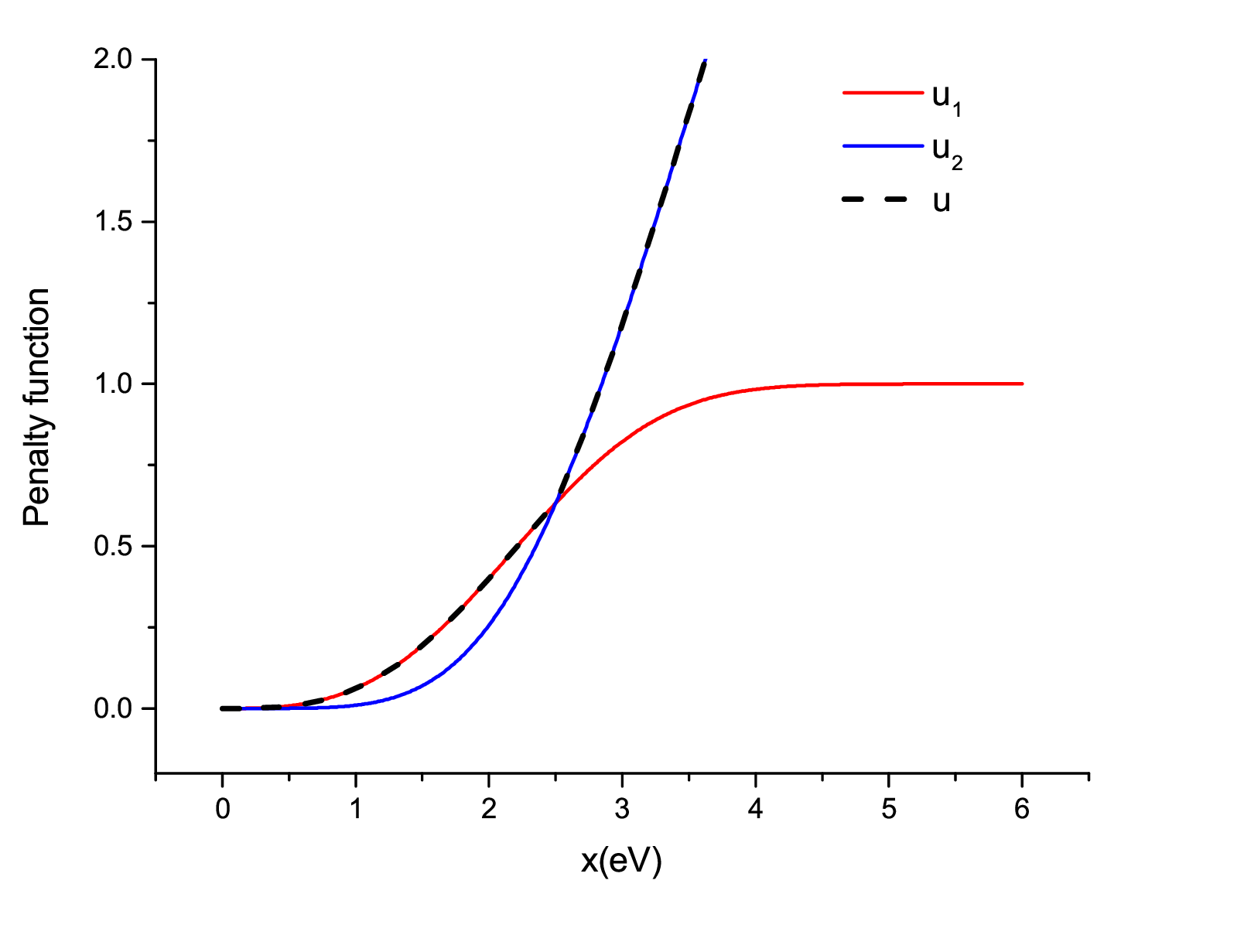} %\par
\par\end{centering}
\caption{Penalty function.
\label{penalty}}
\end{figure}

In the present implementation of LOSC, we define the localized orbitals $\{\phi_i\}$ through a restrained Boys localization procedure.
\begin{align}
    \{\phi_i\} &= \text{argmin} \quad F[\{\phi_i\}; \{\varphi_i\}, \{\epsilon_i^{\rm LO}\}, \{\epsilon_i^{\rm CO}\}]. \label{Boysmin}
\end{align}
Here $\{\varphi_i\}$ and $\{\epsilon_i^{\rm CO}\}$ are auxiliary canonical orbitals and orbital energies of $\bm h_p$ (see the main text); $\{\epsilon_i^{\rm LO}\}$ are introduced as associating to localized orbital energies, which are aligned according to the CO energies, $\epsilon_i^{\rm LO} = \epsilon_i^{\rm CO}$, so that the artificial LO spectrum is identical with the auxiliary CO spectrum. $F$ is the restrained Boys target function, given by
\begin{align}
    F &= \sum_i \Big[\langle \phi_i|\br^2|\phi_i\rangle - \langle \phi_i |\br|\phi_i\rangle ^2\Big] + \sum_{i j} w_{ij} \Big| \langle \phi_i| \varphi_j\rangle \Big|^2 , \label{WBoys}
\end{align}
where the penalty function is a function of the energy difference between an LO and a CO, and we adopt an empirical form as
\begin{equation}
    w_{ij} = w(|\epsilon_i^{\rm LO}-\epsilon_j^{\rm CO}|) = R_0^2 u(|\epsilon_i^{\rm LO}-\epsilon_j^{\rm CO}|)= R_0^2 u(|\epsilon_i^{\rm CO}-\epsilon_j^{\rm CO}|), \label{w}
\end{equation}
where $R_0$ can be considered as related to the typical spread radius of small molecules.
Here $u$ becomes a dimensionless function with the form of (see \Fig{penalty} for the lineshape),
\begin{equation}
u=
\left\{
\begin{array}{cl}
u_1(x), & {\rm if}~x<\epsilon_0,\\
u_2(x),
 & {\rm if}~x\geqslant \epsilon_0,
\end{array}
\right. \label{u}
\end{equation}
where
\begin{equation}
    u_1(x)= \left[1- e^{-\left(\frac{x}{\epsilon_0} \right)^\eta} \right],
\end{equation}
and
\begin{equation}
    u_2(x)=\Big(\frac{x}{\epsilon_0} \Big)^\gamma\, \Big[1- e^{-\left(\frac{x}{\epsilon_0}\right)^\eta} \Big].
\end{equation}
We optimize the parameters for LOSC--BLYP (which are the same as LOSC--LDA, see the next section) to obtain a balanced behavior for thermochemistry, reaction barrier heights and dissociation curves. The resulting parameters are $R_0=2.7$\AA, $\epsilon_0$=2.5eV, $\eta=3.0$ and $\gamma=2.0$, and the plot of penalty function is shown in Figure~\ref{penalty}. In the present version, the LOSC has negligible correction to compact molecules, and therefore has almost no correction to thermochemistry and little correction to transition state species, see Table~\ref{G2-97} for the overall performance of LOSC on thermochemistry and reaction barriers.

\section{Details about curvature formula for GGAs and hybrid functionals}
The LOSC curvature for GGAs have been set to be the same as LDA, which reads
\begin{equation}
\small
    \frac{1}{2}\kappa_{ij}=\frac{1}{2}\iint\frac{\rho_{i}(\br)\rho_{j}(\br')}{|\br-\br'|}d\br d\br'-\frac{1}{3}\tau C_{x}\int[\rho_{i}(\br)]^{2/3}[\rho_{j}(\br)]^{2/3}d\br.
\end{equation}
For global hybrid and range-separated hybrid functionals in general, the Coulomb interaction is separated into attenuated short-range and longe-range,
\begin{equation}
    \frac{1}{r_{12}} = \frac{1-[\alpha + \beta\cdot \text{erf}(\mu r_{12})]}{r_{12}} + \frac{\alpha + \beta\cdot \text{erf}(\mu r_{12})}{r_{12}},
\end{equation}
where $\text{erf}$ is the error function. The first term on the rhs is the attenuated short-range part, treated by GGA exchange, while the second term is the attenuated long-range part, treated by HF exchange. Note the global hybrids correspond to $\beta=0$, while long-range corrected functionals correspond to $\alpha=0$ and $\beta=1$. When both $\alpha$ and $\beta$ are fractional numbers, it corresponds to the most general form. We design the curvature formula for the general hybrid functional as given by
\begin{align}
\frac{1}{2}\kappa_{ij}&=\frac{1}{2}\iint\frac{\rho_{i}(\br)v(|\br-\br'|)\rho_{j}(\br')}{|\br-\br'|}d\br d\br'-\tau(1-\alpha)\frac{1}{3} C_{x}\int[\rho_{i}(\br)]^{2/3}[\rho_{j}(\br)]^{2/3}d\br.  \label{curvature-ij}
\end{align}
where
\begin{equation}
    \frac{v(r_{12})}{r_{12}} = \frac{1-[\alpha + \beta\cdot \text{erf}(\mu r_{12})]}{r_{12}},
\end{equation}
is the attenuated short-range Coulomb interaction.

Note that for the range-separated hybrid functionals, the HOMO--LUMO gaps are usually much larger than LDA and GGAs. To account for this effect, we choose $R_0=2.0$\AA\; for LOSC--CAM--B3LYP, while for LOSC--GGAs and LOSC--B3LYP The parameters are set to be the same as LOSC--LDA.

In terms of $\tau$ parameter, we choose a nonempirical one so that the exchange energy satisfies the following linear condition at half electron:
\begin{equation}
    E_x \big[\frac{1}{2}\,\rho_1 \big] = \frac{1}{2}E_x[\rho_1].
\end{equation}
Note here $\rho_1$ is an arbitrary one electron density. It follows that
\begin{equation}
    -C_x\int \big[\frac{1}{2}\rho_1(\br) \big]^{4/3}d\br - \frac{C_x}{3}\tau \int [\rho_1(\br)]^{4/3}d\br
    \, \Big[\frac{1}{2}-\Big(\frac{1}{2}\Big)^2 \Big] = -\frac{1}{2}C_x\int \big[\rho_1(\br) \big]^{4/3}d\br,
\end{equation}
which leads to
\begin{equation}
    \tau = 6 \Big[1-(\frac{1}{2})^{1/3} \Big] \approx 1.2378.
\end{equation}
Note that the above analysis is based on the frozen orbital assumption. Now suppose one wants to retrieve the right slope produced by the LDA exchange at integer $n=1$, i.e.,
\begin{equation}
    \frac{d }{dn} E_x \big[n\rho_1 \big]\Big|_{n=1^-} = -E_x[\rho_1],
\end{equation}
then it follows
\begin{equation}
    -\frac{4}{3}C_x\int \big[\frac{1}{2}\rho_1(\br) \big]^{4/3}d\br - \frac{C_x}{3}\tau \int [\rho_1(\br)]^{4/3}d\br
    (1-2n)\Big|_{n=1^-} = -\frac{1}{2}C_x\int \big[\rho_1(\br) \big]^{4/3}d\br,
\end{equation}
which leads to $\tau=1$. This is the parameter used in the global scaling correction.

Regarding the numerical implementation, the double integrals in the curvature formula are evaluated using the resolution of identity (or density fitting) technique \cite{Dunlap793396,Vahtras93514}.
\section{Details about the self-consistent implementation of LOSC}
%We restrict to the integer systems.
The LOSC effective Hamiltonian $\Delta \bm h$ is given by
\begin{equation}
     \Delta \bm h = \frac{\delta \Delta E}{\delta \bm \rho_s} = \frac{\delta \Delta E}{\delta \bm \rho_s}\Big|_{\{\phi_i\} }
    +\sum_i \frac{\delta \Delta E}{\delta \phi_i}\Big|_{\bm \rho_s} \frac{\delta \phi_i}{\delta \bm \rho_s}+ \sum_i \frac{\delta \Delta E}{\delta \phi_i^*}\Big|_{\bm \rho_s} \frac{\delta \phi_i^*}{\delta \bm \rho_s},
\end{equation}
where $\Delta E$ is the LOSC correction. Here the first term on the rhs (we denote it as $\Delta \bm h_1$) is the frozen orbital contribution, whereas the last two terms (we denote it as $\Delta \bm h_2$) arise as the orbital relaxation term. $\Delta \bm h_1$ can be written explicitly as
\begin{align}
    \Delta \bm h_1 = \sum_{i} \kappa_{ii}(\frac{1}{2}-\lambda_{ii})| \phi_i\rangle \langle \phi_i|
        -\sum_{i\neq j}\kappa_{ij} \lambda_{ij} | \phi_i\rangle \langle \phi_j|.
\end{align}
On the other hand, $\Delta \bm h_2$ must be evaluated through a multi-step chain rule which leads to a complicated expression.
\begin{comment}
and reads (here we assume everything is real to simplify the expression)
\begin{align}
    \bm h_2&= \sum_{i} \frac{\delta \Delta E}{\delta \phi_i}\Big|_{\bm \rho_s} \frac{\delta \phi_i}{\delta \bm h_p} \frac{\delta \bm h_p}{\delta \bm \rho_s}+\sum_{i} \frac{\delta \Delta E}{\delta \phi_i^*}\Big|_{\bm \rho_s} \frac{\delta \phi_i^*}{\delta \bm h_p} \frac{\delta \bm h_p}{\delta \bm \rho_s}. \label{h2}
\end{align}
Here $\bm h_p$ is the projected effective Hamiltonian (see the main text).
\end{comment}

In the present self-consistent-field (SCF) implementation of LOSC, we adopt the frozen-orbital approximation and ignore the contribution of $\bm \Delta h_2$, and use $\bm h_0+\Delta \bm h \approx \bm h_0+\Delta \bm h_1$ as an approximate effective Hamiltonian to perform the energy minimization through a gradient descent approach with line search.

\section{Supplemental results of LOSC}
In this work, all the numerical results of DFAs and LSC-DFAs were obtained with an in-house developed program QM$^4$D \cite{QM4D}.
In the main text, the basis set used is aug-cc-pVDZ in Fig.~1, 6-311++G(3df, 3pd) in Fig.~2 and Fig.~3(a), cc-pVTZ in Fig.~3(b)-(d) and 6-31++G{**} in Fig.~4. The auxiliary basis set for the density fitting is aug-cc-pVTZ in Figs.~1--3, and DGauss-A2-DFT-Coulomb-fitting basis \cite{DGauss_A2_Coulomb_Fit} in Fig.~4.

All the coupled cluster calculations were done using the Gaussian09 program \cite{g09}. In the main text, the basis set used is 6-31+G{*} for Fig.~4; an even larger basis set would be too demanding with our computational resources.
%\subsection{Dissociation of Z$_2^+$}

\begin{figure}[H]
\begin{centering}
\includegraphics[width=7in]{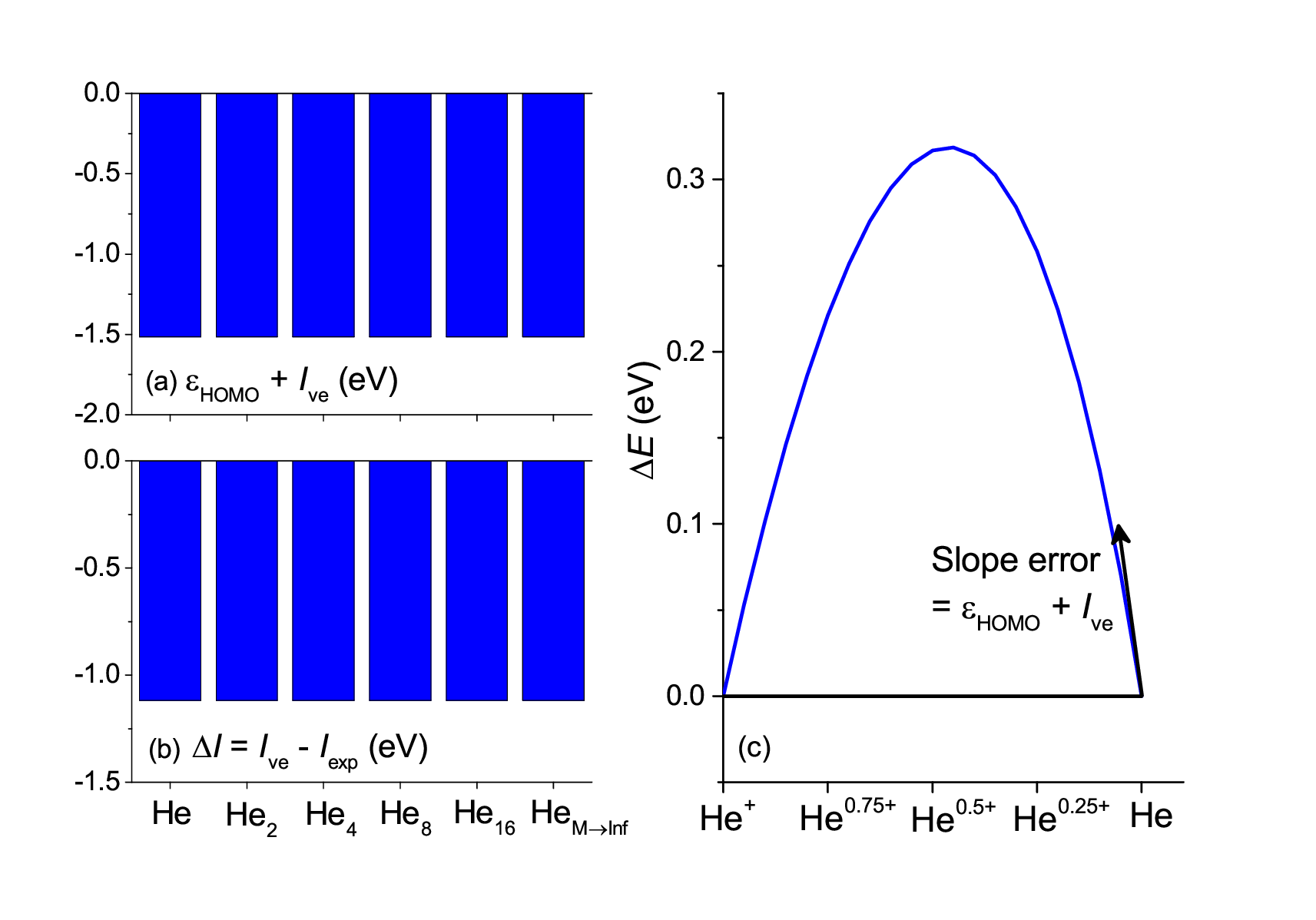} %\par
\par\end{centering}
\caption{ (a) Deviations between the calculated $\ep_{\rm HOMO}$ and $-I_{\rm ve}$ and (b) deviations between $I_{\rm ve}$ and $I_{\rm exp}$ by the HF functional for a series of He$_M$ clusters, as introduced in the main text.
(c) Calculated HF total energy deviation from the linearity condition of a fractionally charged He atom as a function of the fractional charge $\delta$. Here $\Delta E({\rm He}^{\delta+})=E({\rm He}^{\delta+})-\delta E({\rm He}^+)-(1-\delta)E({\rm He})$. \label{HeM_HF}}
\end{figure}

As a supplement to Fig.~1 of the main text, in \Fig{HeM_HF}, we show the error evolution of the HF functional for the loosely bound He$_M$ cluster.
As can be seen in \Fig{HeM_HF}(a) and (b), the deviation of the HOMO energy from the minus integer IP and the deviation of the integer IP from the experimental IP are all independent of $M$, and agree with the error for a single He atom. This is due to the localization error intrinsic in the HF functional, as displayed in \Fig{HeM_HF}(c): its $E(N)$ curve is piecewise concave, which suggests that the fractional system energies are overestimated. Therefore, it tends to avoid forming local fractional charges in He$_M^+$, instead, the removed electron, or the hole, localizes on one of the He atoms, leaving the rest of the He atoms as spectators rather than participants in sharing the hole. This is in contrast to the consequence of the delocalization error as discussed in the main text, and explains why the errors are constants in \Fig{HeM_HF}(a) and (b). In particular, for one He atom, $\ep_{\rm HOMO}+I_{ve}$ agrees with the slope error as indicated by the arrow in \Fig{HeM_HF}(c). It is a negative
error due to the concave nature of the HF $E(N)$ curve. Moreover, we note that the integer IP by HF is not a good approximation to the experimental IP due to the lack of correlation. In particular, the missing correlation energy for the neutral He atom is greater than the He$^+$. As a result, the integer IP is overestimated, as shown in \Fig{HeM_HF}(b) (overestimation translates to negative error). Moreover, the overestimation is in similar magnitude as $\ep_{\rm HOMO}+I_{ve}$, making $\ep_{\rm HOMO}$ a good approximation (slightly underestimated) to $-I_{exp}$ through error cancellation.

As a further remark, for infinite systems, for example the limit of $M\rightarrow \infty$ for the He$_M$ cluster, there are two ways to compute the $E(N)$ curve, i.e., with or without the periodic boundary condition (PBC). Without PBC, the $E(N)$ should be identical with that of a single He atom (concave curve) and the ground state is a symmetry-breaking answer. With PBC, one should obtain a symmetric solution whose energy is higher than without PBC and the $E(N)$ is linear. In the former case, the HOMO energy error relative to the experimental IP transfers to that of a single He atom. In the latter case, however, the HOMO energy error is completely given by the integer IP error of the He$_M$ cluster, which again can be related to the slope error of the $E(N)$ curve of a single He atom, which is identical with the HOMO energy error of a single He atom. This can be deduced by making analogy to the discussion of delocalization error for infinite systems.
Note that all the results of He clusters and discussions presented in this paper on DE or LE for different DFAs in finite systems and bulk systems (periodic or otherwise) are completely consistent with the results of Ref. \cite{Mori-Sanchez08146401}.
In conclusion, the HF HOMO energy error for bulk has close connection with that of a single atom regardless of how the calculation is performed.

\begin{figure}[H]
\begin{centering}
\includegraphics[width=7.8in]{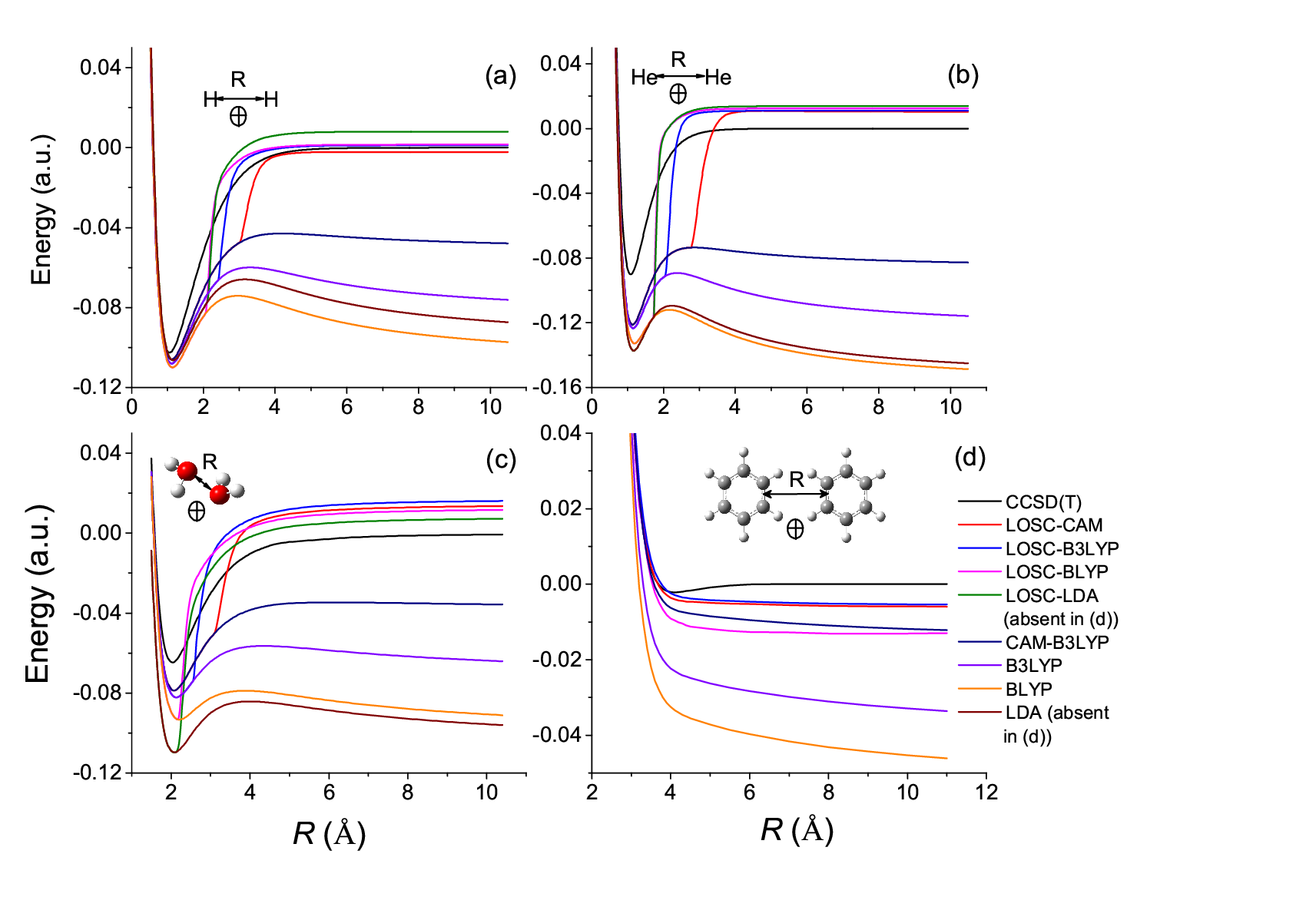} %\par
\par\end{centering}
\caption{Dissociation energy curve of Z$_2^+$, with Z being (a) H atom, (b) He atom, (c) water and (d) benzene. The zero energy is set to be the sum of energies of a Z and Z$^+$. Here the LOSC results are based on post-SCF calculations using the parent functional density.
%The LOSC parameters are $\eta=3.0$, $\epsilon_0=2.5$eV, $\gamma=2.0$, $\tau=1.2378$, $R_0=2.7$\AA\; for LOSC-B3LYP and $R_0=2.0$\AA\; for LOSC-CAMB3LYP.
The basis used for each calculation is (a) 6-311++G(3df,3pd), (b) aug-cc-pVTZ, (c) 6-311++G(3df,3pd) and (d) cc-pVDZ. The fitting basis used for density fitting procedure for the curvature calculation of LOSC is aug-cc-pVTZ for (a)(b)(c), and 6-311++G(3df,3pd) for (d). \label{dissociation}}
\end{figure}

As a supplement to Fig.~2 of the main text, in \Fig{dissociation}, we compare dissociation curves of dimer cations of increasing sizes with the monomer ranging from H, He, water to benzene. Here the results are obtained by doing a one-shot calculation based on the SCF density matrix of the parent DFAs (post-SCF calculation). This is because the parent DFAs give qualitatively right density for these systems and the SCF implementation of LOSC will only slightly change the density. By performing SCF calculation using LOSC-LDA for H$_2^+$ and comparing it with the post-SCF results, the total energy only decreases by no more than 3\,mHartree. In terms of the post-SCF behavior of LOSC-LDA on these four systems, they are qualitatively similar; the dissociation curves are largely improved upon LDA. Moreover, for each dissociation, LOSC on top of different parent functionals lead to similar results and they all improve tremendously over their parent DFAs.

\begin{figure}[H]
\begin{centering}
\includegraphics[width=6.5in]{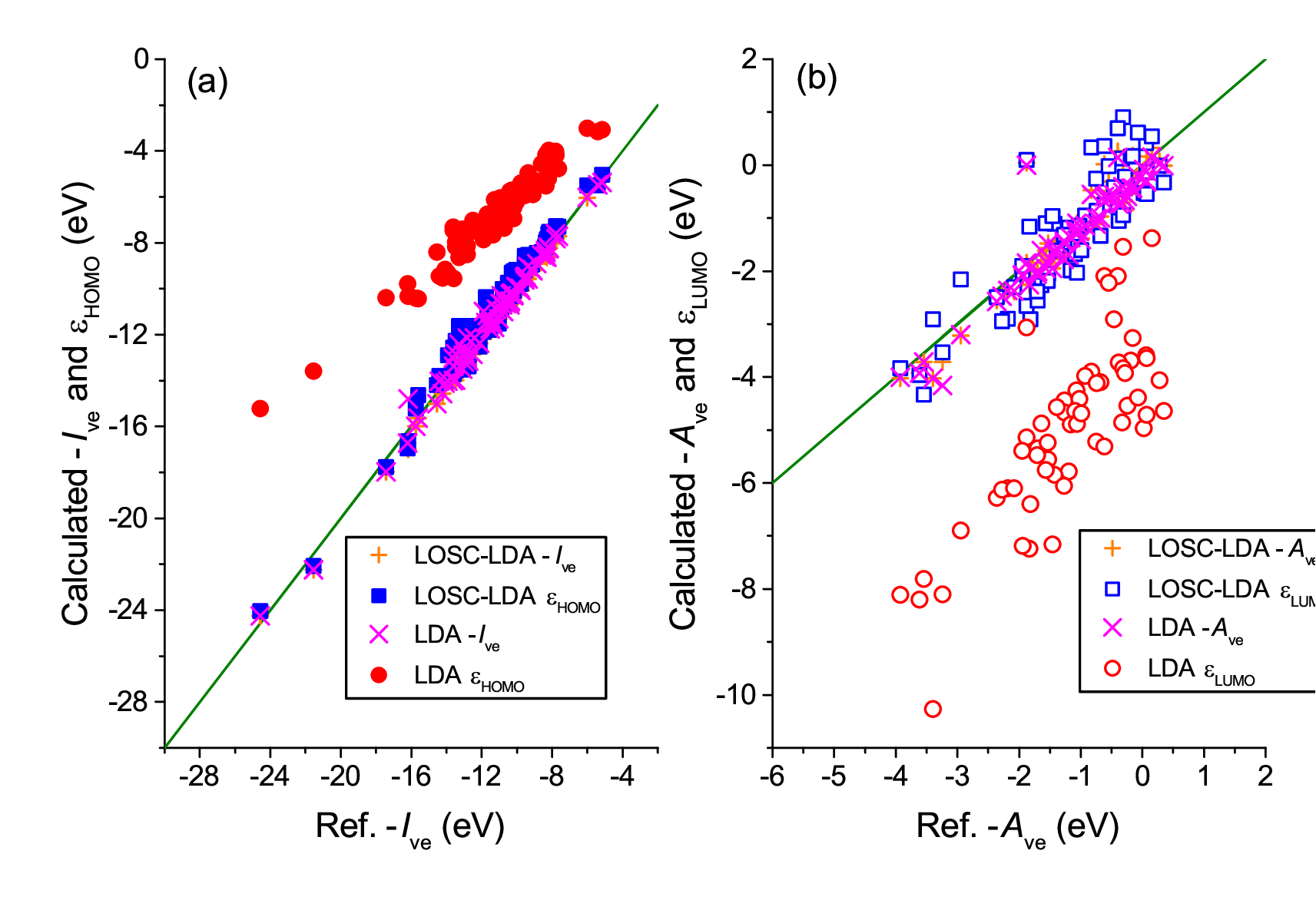} %\par
\par\end{centering}
\caption{(a) Calculated $-I_{\rm ve}^{\rm DFA}$ and $\epsilon_{\rm HOMO}$ of LDA and LOSC-LDA in comparison with the reference $-I_{\rm ve}$ for 82 atoms and molecules of the modified G2--97 set \cite{Curtiss971063, footnote1}; (b) Calculated $-A_{\rm ve}^{\rm DFA}$ and $\epsilon_{\rm LUMO}$ of LDA and LOSC-LDA in comparison with the reference $-A_{\rm ve}$ for 61 atoms and molecules of the modified G2--97 set.
The results of $-I_{\rm ve}$ ($-A_{\rm ve}$) presented here are supplemental to the results of $\epsilon_{\rm HOMO}$ ($\epsilon_{\rm LUMO}$) presented in Figure 3(a) of the main text.
Here the LOSC-LDA $-I_{\rm ve}$ ($-A_{\rm ve}$) results almost overlap with LDA.
The basis used for each calculation is 6-311++G(3df,3pd) and the fitting basis used for density fitting procedure in the curvature calculation of LOSC is aug-cc-pVTZ. \label{IP-EA}}
\end{figure}

As a supplement to Fig.~3(a) of the main text, in Figure~\ref{IP-EA}, we compare $-I_{\rm ve}^{\rm DFA}$ ($-A_{\rm ve}^{\rm DFA}$) of LDA and LOSC-LDA (along with their $\epsilon_{\rm HOMO}$, $\epsilon_{\rm LUMO}$) with the best estimates of the vertical $-I_{\rm ve}$ ($-A_{\rm ve}$) obtained from experiment for atoms and CCSD(T) calculation (with basis set extrapolation) for molecules.
As can be seen, the LOSC-LDA predicted integer IP (EA) are in good agreement with LDA and the error relative to the best estimates is rather small, which verifies that LOSC has little effect on the total energy for compact molecules.
Moreover, for LOSC-LDA, HOMO (LUMO) energies reach better agreement with $-I_{\rm ve}^{\rm LOSC-LDA}$ ($-A_{\rm ve}^{\rm LOSC-LDA}$).

\begin{figure}[H]
\begin{centering}
\includegraphics[width=6.5in]{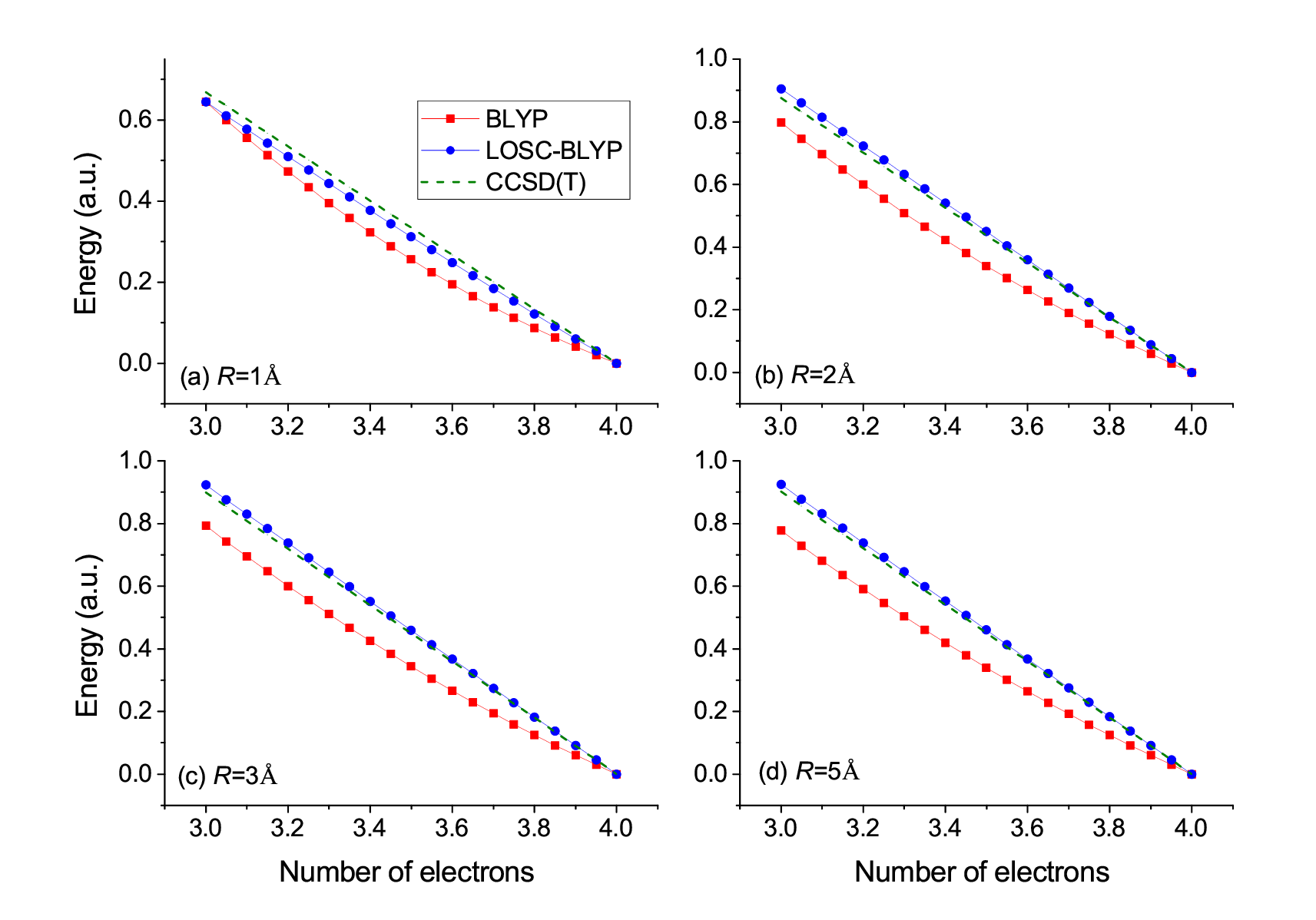} %\par
\par\end{centering}
\caption{$E$ vs $N$ curve regarding fractional electron removal of He$_2$ with different internuclear distance. Here the energy of He$_2$ is set to zero for each method in all the subplots. The CCSD(T) results are obtained by connecting the integer energies with dashed straight lines.
Here the LOSC results are based on post-SCF calculations using the parent functional reduced density matrix.
The basis used for all calculations is aug-cc-pVTZ. The fitting basis used for the density fitting procedure in the curvature calculation of LOSC is aug-cc-pVTZ. \label{EvsN}}
\end{figure}

\begin{figure}[H]
\begin{centering}
\includegraphics[width=6.5in]{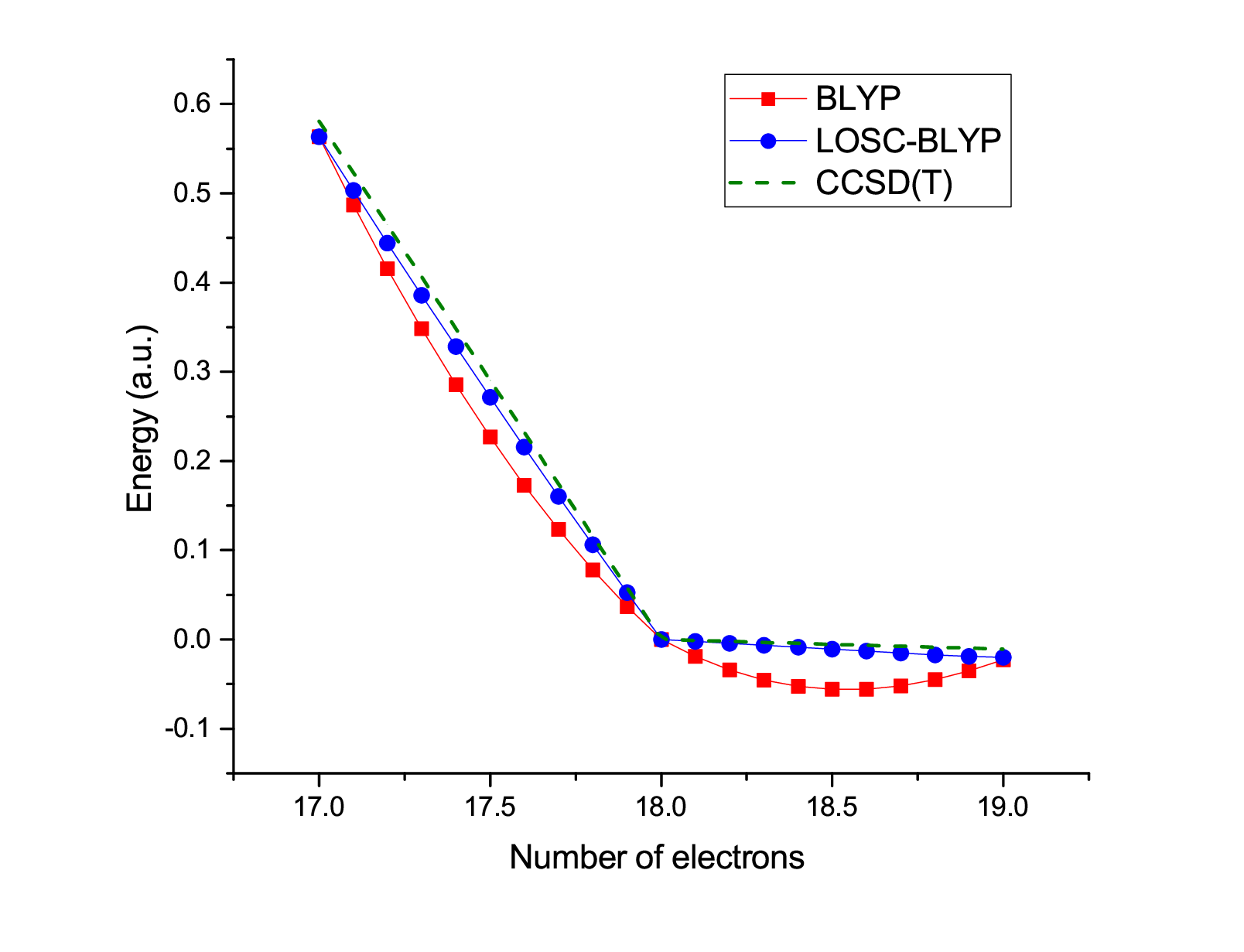} %\par
\par\end{centering}
\caption{$E$ vs $N$ curve of F$_2$ with bond length $R=1.412$\AA. Here the energy of F$_2$ is set to zero for each method The CCSD(T) results are obtained by connecting the integer energies with dashed straight lines.
Here the LOSC results are based on post-SCF calculations using the parent functional reduced density matrix.
The basis used for all calculations is aug-cc-pVTZ. The fitting basis used for the density fitting procedure in the curvature calculation of LOSC is aug-cc-pVTZ. \label{EvsN_F2}}
\end{figure}

Furthermore, to validate the linearity condition for fractional charge systems, we compare the $E$ vs $N$ curve regarding electron removal of He$_2$ with increasing internuclear distances for BLYP and LOSC-BLYP, benchmarked by CCSD(T) as shown in Figure~\ref{EvsN}. As can be seen, the BLYP exhibits convex curves for all $R$, while the correct linear curve is largely restored by LOSC-BLYP. It is worth noticing that the LOSC not only changes the wrong curvature, but also corrects the wrong integer point. For example, when $R=3$ and 5\AA, the $N-1$ system energy is greatly underestimated by BLYP, but reaches good agreement with the CCSD(T) result when applying LOSC. In Figure~\ref{EvsN_F2}, we show the $E$ vs $N$ curve for F$_2$ molecule, in the case of electron removal and electron addition. As can be seen, both curves are convex for BLYP but straightened by LOSC-BLYP.

In Table~\ref{LOSC-gap-1}, we further present the LOSC results in terms of mean absolute deviation of $-I_{\rm ve}^{\rm DFA}$ ($-A_{\rm ve}^{\rm DFA}$) and $\epsilon_{\rm HOMO}$ ($\epsilon_{\rm LUMO}$) based on LDA, GGAs, global hybrid and range separated functionals, relative to the best estimates of the $-I_{\rm ve}$ ($-A_{\rm ve}$).
%
%Here, as in the main text, the calculated HOMO, LUMO of DFAs or LOSC-DFAs are compared with the vertical -IP and -EA of DFAs or LOSC-DFAs.
Here, as in the main text, the calculations are done in the post-SCF manner, in the sense that the SCF converged density of the parent functional is used to evaluated $\bm h_0+ \Delta \bm h_1$ and $\Delta E$. And the LOSC orbital energy is computed through
\begin{equation}
    \epsilon_i^{\rm LOSC-DFA} = \langle \psi_i | \bm h_0 + \Delta \bm h_1|\psi_i \rangle,
\end{equation}
where $\psi_i$'s are the canonical orbitals of the parent functional.

In fact, an alternative way of computing the orbital energy is through diagonalizing the LOSC effective Hamiltonian $\bm h_0 + \Delta \bm h_1$ in the post-SCF manner. This is in the similar spirit as the SCF implementation, but in a one shot calculation. We have compared the orbital energy results of LOSC-LDA using post-SCF projection and post-SCF diagonlization, and they differ by 0.02eV at most for atoms or molecules in the G2-97 set.

Regarding Fig.~3(b)-(d) of the main text, we have truncated the number of LOs in the localization and procedures following that to save computational cost without affecting the numerical accuracy. Essentially, the orbitals (COs or LOs) with too low or too high energies have little contribution to the total energy or orbital energies near the frontier level, although they do affect the correction to the lower-lying or higher-lying orbital energies.
Therefore, for the correct prediction of HOMO/LUMO energies and photoemission spectra near the frontier level, we have restricted the energy range from -30\,eV to 10\,eV and only consider the LOs within this range to tremendously reduce the computational cost during the localization and curvature computation procedures. By comparing the results with and without truncation for polyacenes with $n=1$--3, we find that energy levels of interest change by less than 0.01\,eV for LOSC-LDA. Here the calculations have been based on post-SCF projection.
In Table~\ref{acene-gap}, we supplement the LOSC results of polyacenes using other mainstream DFAs as parent functionals. The global scaling correction (GSC) results are also listed as a comparison. From Table\,\ref{acene-gap}, it is obvious that the effectiveness of LOSC is insensitive to the parent DFAs; and its behavior is rather size-independent. This is in clear contrast to the GSC, which behaves very well for small-sized systems, but becomes insufficient as the system size increases.

Finally, the LOSC largely preserves the thermochemistry of the parent functionals, as can be seen from Table~\ref{G2-97}. For reaction barrier heights, the LOSC results are essentially similar to the parent functional performance, despite slight improvement of LOSC-BLYP over BLYP. This is because for compact molecules at equilibrium geometries, for which the thermochemical properties are calculated, the restrained Boys localization takes little effect (almost no orbital mixing), rendering $\phi_i \approx \varphi_i$. Moreover, $\varphi_i$'s are eigen-orbitals of the projected Hamiltonian $\bm h_p$, so that the set $\{\varphi_i\}$ can be clearly separated into two subsets, spanning the occupied and virtual space, respectively; and the set $\{\phi_i\}$ can be nearly separated into the two subsets. As a result, the diagonal elements of the local occupation matrix $\lambda_{ii} = \langle \phi_i | \rho | \phi_i \rangle \approx$ 0 or 1, and the off-diagonal elements $\lambda_{ij} = \langle \phi_i | \rho | \phi_j \rangle \approx 0$. Then it is obvious that the correction energy is small, hence the thermochemistry is preserved. When the bond is stretched, corresponding to transition state species, then the restrained Boys localization takes effect and orbitals from occupied and virtual spaces begin to mix into $\phi_i$'s, yielding nonzero $\lambda_{ii}$ and $\lambda_{ij}$. This is when we see the correction in reaction barriers and dissociating molecules. In the present paper, however, we adopt a conservative localization scheme, so that slight stretch of molecules does not incur orbital mixing from occupied and virtual space. This leads to little correction to the reaction barrier heights.

\begin{comment}
\begin{table}[H]
{\caption{{LOSC overall performance (based on post-SCF calculation) in terms of mean absolute error for HOMO and LUMO in comparison with the integer $-I_{\rm ve}$ and $-A_{\rm ve}$, respectively. Basis used is 6-311++G(3df,3pd). Fitting basis: aug-cc-pVTZ.
\label{LOSC-gap}}}
}
\centering
\begin{tabular}{cccccccccc}
\hline
\multirow{2}{*}{} & \multirow{2}{*}{} & \multirow{2}{*}{LDA} & LOSC- & \multirow{2}{*}{BLYP} & LOSC- & \multirow{2}{*}{B3LYP} & LOSC- & \multirow{2}{*}{CAM-B3LYP} & \multicolumn{1}{c}{LOSC-}\tabularnewline
 &  &  & LDA &  & BLYP &  & B3LYP &  & CAM\tabularnewline
\hline
\multirow{2}{*}{Atoms} & IP (eV) & 5.07 & 0.50 & 5.06 & 0.47 & 3.88 & 0.37 & 2.27 & 0.46\tabularnewline
 & EA (eV) & 3.19 & 0.61 & 3.06 & 0.63 & 2.28 & 0.62 & 1.11 & 0.34\tabularnewline
\hline
\multirow{2}{*}{Molecules} & IP (eV) & 4.19 & 0.47 & 4.11 & 0.43 & 3.07 & 0.32 & 1.46 & 0.33\tabularnewline
 & EA (eV) & 3.65 & 0.43 & 3.52 & 0.43 & 2.58 & 0.36 & 1.12 & 0.27\tabularnewline
\hline
\end{tabular}
\end{table}
\end{comment}

\setlength{\tabcolsep}{9pt}
\setlength{\extrarowheight}{2pt}

\begin{table}[H]
{\caption{{
LOSC overall performance (based on post-SCF calculation) in terms of mean absolute error for $-I_{\rm ve}$, $-A_{\rm ve}$, HOMO and LUMO energies in comparison with the best estimate of $-I_{\rm ve}$ and $-A_{\rm ve}$, respectively. All energies are in unit of eV. Here the basis used is 6-311++G(3df,3pd) and the auxiliary basis for density fitting is aug-cc-pVTZ.
\label{LOSC-gap-1}}}
}
\centering
\begin{tabular}{ccccccccc}
\hline
\multirow{2}{*}{} & \multirow{2}{*}{LDA} & LOSC- & \multirow{2}{*}{BLYP} & LOSC- & \multirow{2}{*}{B3LYP} & LOSC- & \multirow{2}{*}{CAM} & LOSC-\tabularnewline
 &  & LDA &  & BLYP &  & B3LYP &  & CAM\tabularnewline
\hline
$-I_{\rm ve}$ & 0.26 & 0.24 & 0.32 & 0.27 & 0.19 & 0.18 & 0.17 & 0.17\tabularnewline
$-A_{\rm ve}$ & 0.28 & 0.28 & 0.16 & 0.23 & 0.16 & 0.18 & 0.20 & 0.20\tabularnewline
$\epsilon_{\rm HOMO}$ & 4.43 & 0.50 & 4.61 & 0.63 & 3.31 & 0.35 & 1.66 & 0.37\tabularnewline
$\epsilon_{\rm LUMO}$ & 3.76 & 0.50 & 3.34 & 0.52 & 2.55 & 0.48 & 1.18 & 0.39\tabularnewline
\hline
\end{tabular}
\end{table}

\setlength{\tabcolsep}{1.8pt}
\setlength{\extrarowheight}{1.5pt}

\begin{table}[H]
{\caption{{Calculated HOMO, LUMO and HOMO-LUMO gap in comparison with experimental value for polyacenes. All energies are in unit of eV. Here the basis used is cc-pVTZ and the auxiliary basis for density fitting is aug-cc-pVTZ.
\label{acene-gap}}}
}
\centering
\begin{tabular}{cccccccccccccc}
\hline
\multirow{2}{*}{} & \multirow{2}{*}{} & \multirow{2}{*}{Exp} & \multirow{2}{*}{LDA} & LOSC- & \multirow{2}{*}{SLDA} & \multirow{2}{*}{PBE} & LOSC- & \multirow{2}{*}{BLYP} & LOSC- & \multirow{2}{*}{B3LYP} & LOSC- & \multirow{2}{*}{CAM} & LOSC-\tabularnewline
 &  &  &  & LDA &  &  & PBE &  & BLYP &  & B3LYP &  & CAM\tabularnewline
\hline
\multirow{7}{*}{HOMO} & 1 & -9.24 & -6.50 & -8.93 & -9.28 & -6.28 & -8.69 & -6.09 & -8.51 & -7.04 & -8.96 & -8.47 & -8.98\tabularnewline
 & 2 & -8.11 & -5.68 & -8.16 & -7.63 & -5.46 & -8.01 & -5.26 & -7.71 & -6.11 & -8.05 & -7.41 & -8.00\tabularnewline
 & 3 & -7.47 & -5.19 & -7.66 & -6.75 & -4.97 & -7.45 & -4.76 & -7.23 & -5.54 & -7.36 & -6.74 & -7.30\tabularnewline
 & 4 & -6.97 & -4.87 & -7.21 & -6.20 & -4.64 & -7.00 & -4.44 & -6.80 & -5.16 & -6.89 & -6.29 & -6.78\tabularnewline
 & 5 & -6.63 & -4.66 & -6.82 & -5.81 & -4.42 & -6.59 & -4.22 & -6.42 & -4.90 & -6.46 & -5.97 & -6.28\tabularnewline
 & 6 & -6.36 & -4.50 & -6.62 & -5.50 & -4.26 & -6.40 & -4.06 & -6.15 & -4.71 & -6.24 & -5.73 & -6.08\tabularnewline
\cline{2-14}
 & MAE & - & 2.23 & 0.21 & 0.62 & 2.46 & 0.13 & 2.66 & 0.33 & 1.89 & 0.14 & 0.70 & 0.23\tabularnewline
\hline
\multirow{7}{*}{LUMO} & 1 & 1.12 & -1.38 & 0.71 & 1.02 & -1.13 & 0.96 & -0.97 & 1.12 & -0.39 & 1.25 & 0.90 & 1.22\tabularnewline
 & 2 & -0.2 & -2.25 & -0.16 & -0.41 & -2.00 & 0.09 & -1.82 & 0.26 & -1.33 & 0.31 & -0.14 & 0.18\tabularnewline
 & 3 & -0.53 & -2.83 & -0.77 & -1.20 & -2.58 & -0.48 & -2.39 & -0.33 & -1.97 & -0.37 & -0.87 & -0.55\tabularnewline
 & 4 & -1.06 & -3.22 & -1.26 & -1.90 & -2.96 & -1.00 & -2.76 & -0.78 & -2.41 & -0.94 & -1.37 & -1.08\tabularnewline
 & 5 & -1.39 & -3.48 & -1.59 & -2.32 & -3.23 & -1.32 & -3.02 & -1.13 & -2.71 & -1.38 & -1.74 & -1.49\tabularnewline
 & 6 & -1.466 & -3.67 & -1.77 & -2.62 & -3.42 & -1.48 & -3.21 & -1.32 & -2.93 & -1.52 & -2.00 & -1.76\tabularnewline
\cline{2-14}
 & MAE & - & 2.22 & 0.23 & 0.65 & 1.97 & 0.11 & 1.77 & 0.22 & 1.37 & 0.16 & 0.30 & 0.15\tabularnewline
\hline
\multirow{7}{*}{Gap} & 1 & 10.36 & 5.12 & 9.64 & 10.30 & 5.15 & 9.65 & 5.12 & 9.63 & 6.65 & 10.21 & 9.37 & 10.20\tabularnewline
 & 2 & 7.91 & 3.43 & 8.00 & 7.22 & 3.46 & 8.10 & 3.44 & 7.97 & 4.78 & 8.36 & 7.27 & 8.18\tabularnewline
 & 3 & 6.94 & 2.36 & 6.89 & 5.55 & 2.39 & 6.97 & 2.37 & 6.90 & 3.57 & 6.99 & 5.87 & 6.75\tabularnewline
 & 4 & 5.91 & 1.65 & 5.95 & 4.30 & 1.68 & 6.00 & 1.68 & 6.02 & 2.75 & 5.95 & 4.92 & 5.70\tabularnewline
 & 5 & 5.24 & 1.18 & 5.23 & 3.49 & 1.19 & 5.27 & 1.20 & 5.29 & 2.19 & 5.08 & 4.23 & 4.79\tabularnewline
 & 6 & 4.91 & 0.83 & 4.85 & 2.88 & 0.84 & 4.92 & 0.85 & 4.83 & 1.78 & 4.72 & 3.73 & 4.32\tabularnewline
\cline{2-14}
 & MAE & - & 4.45 & 0.16 & 1.25 & 4.42 & 0.18 & 4.43 & 0.18 & 3.26 & 0.17 & 0.98 & 0.31\tabularnewline
\hline
\end{tabular}
\end{table}

\setlength{\tabcolsep}{5pt}
\setlength{\extrarowheight}{2pt}

\begin{table}[H]
{\caption{{LOSC overall performance (based on post-SCF calculation) on reaction barriers and atomization energies for G2-97 set. All energies are in unit of kcal/mol. Here the basis used is 6-311++G(3df,3pd) and the auxiliary basis for density fitting is aug-cc-pVTZ.
\label{G2-97}}}
}
\centering
\footnotesize
\begin{tabular}{cccccccc}
\hline
\multirow{2}{*}{} & \multirow{2}{*}{} & \multirow{2}{*}{BLYP} & LOSC- & \multirow{2}{*}{B3LYP} & LOSC- & \multirow{2}{*}{CAM-B3LYP} & LOSC-\tabularnewline
 &  &  & BLYP &  & B3LYP &  & CAM\tabularnewline
\hline
\multirow{2}{*}{BH} & HTBH & 7.83 & 7.02 & 4.43 & 4.40 & 3.29 & 3.29\tabularnewline
 & NHTBH & 8.51 & 7.66 & 4.44 & 4.40 & 2.67 & 2.67\tabularnewline
\hline
\multirow{6}{*}{G2-97} & G2-1 (55) & 4.98 & 4.95 & 2.70 & 2.79 & 3.28 & 3.28\tabularnewline
 & Non-hydrogen (21) & 14.06 & 13.94 & 7.41 & 7.42 & 6.05 & 6.06\tabularnewline
 & Hydrocarbons (17) & 10.35 & 10.44 & 3.45 & 3.48 & 2.53 & 2.52\tabularnewline
 & Substituted Hydrocarbons (42) & 6.23 & 6.27 & 2.52 & 2.54 & 3.59 & 3.58\tabularnewline
 & Inorganic hydrides and radicals (13) & 5.49 & 5.54 & 2.17 & 2.15 & 3.91 & 3.90\tabularnewline
 & G2 mean & 7.28 & 7.28 & 3.36 & 3.40 & 3.73 & 3.73\tabularnewline
\hline
\end{tabular}
\end{table}